\documentclass[journal]{IEEEtran}
\usepackage[noadjust]{cite}
\usepackage{amsmath,amssymb,amsfonts}
\usepackage{algorithmic}
\usepackage{graphicx}
\usepackage{textcomp}
\usepackage{xcolor}
\usepackage{hhline}
\usepackage{multirow}

\usepackage{amsmath,graphicx}

\usepackage[british]{babel}
\usepackage{amsfonts}
\usepackage{amssymb}
\usepackage{graphicx}
\usepackage[export]{adjustbox}
\usepackage{mathtools}
\usepackage{siunitx}
\usepackage{makecell}
\usepackage[pagebackref=true,colorlinks=false,breaklinks=true,letterpaper=true,bookmarks=false]{hyperref} 
\usepackage{algorithmic}
\usepackage{ctable} 
\usepackage{gensymb}

\newcommand{\etal}{\textit{et al}.~}
\newcommand{\ie}{\textit{i}.\textit{e}.~}

\newcommand\norm[1]{\left\lVert#1\right\rVert}

\newcommand{\figref}[1]{Fig.\,\ref{#1}}
\newcommand{\tabref}[1]{Table\,\ref{#1}}
\newcommand{\secref}[1]{Section\,\ref{#1}}

\renewcommand\thetable{\Roman{table}}

\newcommand{\RNum}[1]{\uppercase\expandafter{\romannumeral #1\relax}}

\def\wrt{\textit{w.r.t.~}}

\usepackage{array}

\ifCLASSOPTIONcompsoc
  \usepackage[caption=false,font=normalsize,labelfont=sf,textfont=sf]{subfig}
\else
  \usepackage[caption=false,font=footnotesize]{subfig}
\fi

\usepackage{url}

\begin{document}
%
% paper title
% Titles are generally capitalized except for words such as a, an, and, as,
% at, but, by, for, in, nor, of, on, or, the, to and up, which are usually
% not capitalized unless they are the first or last word of the title.
% Linebreaks \\ can be used within to get better formatting as desired.
% Do not put math or special symbols in the title.
%\title{Deep Residual Network for Shearlet-Regularized Light Field Reconstruction}
\title{DRST: Deep Residual Shearlet Transform for Densely-Sampled Light Field Reconstruction}
%
%
% author names and IEEE memberships
% note positions of commas and nonbreaking spaces ( ~ ) LaTeX will not break
% a structure at a ~ so this keeps an author's name from being broken across
% two lines.
% use \thanks{} to gain access to the first footnote area
% a separate \thanks must be used for each paragraph as LaTeX2e's \thanks
% was not built to handle multiple paragraphs
%

\author{Yuan~Gao, %~\IEEEmembership{Student Member,~IEEE,}
        Robert~Bregovi\'{c},~\IEEEmembership{Member,~IEEE,}
        Reinhard~Koch,~\IEEEmembership{Member,~IEEE,} and
        Atanas~Gotchev,~\IEEEmembership{Member,~IEEE}% <-this % stops a space
\thanks{Y. Gao, R. Bregovi\'{c} and A. Gotchev are with the Faculty of Information Technology and Communication Sciences (ITC), Tampere University, 33014 Tampere, Finland. 
(e-mail: \{\href{mailto:yuan.gao@tuni.fi}{yuan.gao}, \href{mailto:robert.bregovic@tuni.fi}{robert.bregovic}, \href{mailto:atanas.gotchev@tuni.fi}{atanas.gotchev}\}@tuni.fi)}% <-this % stops a space
\thanks{R. Koch is with the Department of Computer Science, Kiel University, 24118 Kiel, Germany. 
(e-mail: \href{mailto:rk@informatik.uni-kiel.de}{rk}@informatik.uni-kiel.de)}% <-this % stops a space
\thanks{The work in this paper was funded from the European Union's Horizon 2020 research and innovation program under the Marie Sk\l{}odowska-Curie grant agreement No.\,676401, European Training Network on Full Parallax Imaging, and the German Research Foundation (DFG) No.\,K02044/8-1. The authors thank Nvidia for GPU hardware donations.} % The Titan Xp used for this research was donated by the NVIDIA Corporation.}
%\thanks{Manuscript received April 19, 2005; revised August 26, 2015.}
}

\maketitle

% As a general rule, do not put math, special symbols or citations
% in the abstract or keywords.
\begin{abstract}
The Image-Based Rendering (IBR) approach using Shearlet Transform (ST) is one of the most effective methods for Densely-Sampled Light Field (DSLF) reconstruction. 
The ST-based DSLF reconstruction typically relies on an iterative thresholding algorithm for Epipolar-Plane Image (EPI) sparse regularization in shearlet domain, involving dozens of transformations between image domain and shearlet domain, 
which are in general time-consuming. 
To overcome this limitation, a novel learning-based ST approach, referred to as Deep Residual Shearlet Transform (DRST), is proposed in this paper. 
Specifically, for an input sparsely-sampled EPI, 
DRST employs a deep fully Convolutional Neural Network (CNN) to predict the residuals of the shearlet coefficients in shearlet domain in order to reconstruct a densely-sampled EPI in image domain. 
The DRST network is trained on synthetic Sparsely-Sampled Light Field (SSLF) data only by leveraging elaborately-designed masks. 
Experimental results on three challenging real-world light field evaluation datasets with varying moderate disparity ranges
(8\,-\,16 pixels) demonstrate the superiority of the proposed learning-based DRST approach over the non-learning-based ST method for DSLF reconstruction. 
Moreover, DRST provides a 2.4x speedup over ST, at least.
\end{abstract}

% Note that keywords are not normally used for peerreview papers.

% For peer review papers, you can put extra information on the cover
% page as needed:
% \ifCLASSOPTIONpeerreview
% \begin{center} \bfseries EDICS Category: 3-BBND \end{center}
% \fi
%
% For peerreview papers, this IEEEtran command inserts a page break and
% creates the second title. It will be ignored for other modes.
\IEEEpeerreviewmaketitle

%\vspace{-.6em}
\section{Introduction}
\label{sec:intro}
\IEEEPARstart{D}{ensely-Sampled} Light Field (DSLF) is a discrete representation of the 4D approximation of the plenoptic function parameterized by two parallel planes (camera plane and image plane)
\cite{levoy1996light}, 
where multi-perspective camera views are arranged in such a way that the disparity ranges between adjacent views are less than or equal to one pixel
\cite{vagharshakyan2015image}.
DSLF has a wide range of applications, such as depth estimation, super-resolution and synthetic aperture imaging   
\cite{wu2017light}, 
visualization on 
3DTV
\cite{smolic20113d}
and 
Virtual Reality (VR)
\cite{yu2017light}
devices. 
In real-world environments, a DSLF is extremely difficult to capture by modern light field acquisition systems, 
such as  
micro-lens array (MLA)
\cite{ng2005light,perwass2012single},
multi-camera array 
\cite{wilburn2005high,flynn2019deepview,lu2019high} 
and coded mask
\cite{babacan2012compressive,marwah2013compressive}. 
Nevertheless, these state-of-the-art light field devices are successful in capturing Sparsely-Sampled Light Fields (SSLFs), where the disparity ranges of  any two neighboring views are larger than one pixel.
%the light fields without satisfying the fundamental disparity requirement of DSLF can generally be captured by such light field acquisition systems and called Sparsely-Sampled Light Fields (SSLFs). 
Therefore, for real-world scenes, DSLFs are typically reconstructed from SSLFs.
This paper studies how to effectively and efficiently reconstruct a DSLF for a real-world SSLF.

The Shearlet Transform (ST)-based DSLF reconstruction algorithm %-based light field reconstruction algorithm 
\cite{vagharshakyan2018light,vagharshakyan2017accelerated}
is one of the state-of-the-art Image-Based Rendering (IBR) approaches
\cite{shum2008image,shum2003survey}, 
which treats an input SSLF as a set of sparsely-sampled Epipolar-Plane Images (EPIs) and leverages the sparse representation of these EPIs in shearlet domain to perform densely-sampled EPI reconstruction in image domain. 
However, the sparse regularization by ST is an iterative algorithm that involves dozens of iterations of domain transformations, \ie shearlet analysis transform from image domain to shearlet domain and shearlet synthesis transform from shearlet domain to image domain. 
To be more precise, a shearlet analysis transform converts an input grayscale EPI into $\eta$ shearlet coefficients, 
which requires one 2D Discrete Fourier Transform (DFT) and $\eta$ 2D inverse DFTs.
On the contrary, a shearlet synthesis transform converts the regularized $\eta$ shearlet coefficients into the output grayscale EPI, 
requiring $\eta$ 2D DFTs and one 2D inverse DFT. 
As a result, ST tends to be time-consuming for DSLF reconstruction on SSLFs with large spatial or large angular resolution. 

To address this fundamental issue, a novel learning-based approach, referred to as Deep Residual Shearlet Transform (DRST), is proposed in this paper. 
In particular, DRST performs shearlet coefficient reconstruction in shearlet domain for an input sparsely-sampled EPI by means of a deep Convolutional Neural Network (CNN), which is composed of a residual learning strategy and an encoder-decoder network that predicts the residuals of the shearlet coefficients.
The reconstructed shearlet coefficients in shearlet domain are then transformed back into image domain to produce a corresponding inpainted densely-sampled EPI. 
Finally, a target DSLF can be reconstructed by repeating this EPI reconstruction process on all the sparsely-sampled EPIs of the input SSLF.
Besides, the network of DRST is fully convolutional and end-to-end trainable.
Considering the aforementioned difficulty of acquiring ground-truth DSLFs, the training of DRST is performed on SSLF data only.
The synthetic SSLF data are used for training because the ground-truth disparity information,
which is important to the shearlet system construction, pre- and post-shearing steps of DRST,
can be provided by using the state-of-the-art 3D computer graphics softwares.

The key contributions of this paper are as follows.
\begin{itemize}
\item 
We propose a learning-based DRST method that achieves better DSLF reconstruction performance than the non-learning-based ST algorithm on three evaluation datasets composed of real-world  horizontal-parallax light fields with different moderate disparity ranges (8\,-\,16 pixels);
\item 
The network of DRST is trained on synthetic SSLF data by means of the elaborately-designed masks. 
To our best knowledge, this is the first work to investigate learning-based DSLF reconstruction with only exploiting synthetic SSLFs as training data; 
\item
The proposed learning-based DRST is more time-efficient than the non-learning-based ST. Specifically, DRST provides a 2.4x speedup over ST, at least. 
\end{itemize}

%mainly because DRST requires only one iteration of shearlet domain transformations. 

The paper is organized as follows. 
\secref{sec:related} first introduces the related work on DSLF reconstruction and then outlines how to employ the non-learning-based ST for DSLF reconstruction. 
In \secref{sec:method}, we detail the proposed learning-based DRST.
\secref{sec:experiment} is devoted to the experiments and analysis of DRST and other baseline approaches. 
Finally, \secref{sec:conclusion} concludes and summarizes this paper.

%\vspace{-.6em}
\section{Related Work} \label{sec:related}
As pointed out in the introduction to this paper, 
%As explained in the previous section, 
the modern light field acquisition systems can hardly capture DSLFs in real-world environments due to their hardware limitations; 
however, 
a real-world SSLF with a moderate disparity range (8\,-\,16 pixels) is possible to capture by most of them.  
%a real-world DSLF can hardly be captured by them, while a real-world SSLF with a moderate disparity range ($8$\,-\,$16$ pixels) can easily be captured by most of these systems. 
Therefore, performing an effective and efficient DSLF reconstruction on the captured SSLFs with moderate disparity ranges is the best way to compensate for the hardware limitations of these modern light field acquisition systems. 
The DSLF reconstruction problem can potentially be solved by several approaches that are categorized into two types, \ie learning-based novel view synthesis and light field angular super-resolution. 
Regarding the former type, 
Niklaus \etal propose a spatially-adaptive Separable Convolution (SepConv) approach that employs a CNN  to predict the separable 1D kernels for video frame synthesis
\cite{niklaus2017iccv}.
Gao and Koch propose a fine-tuning strategy for SepConv, referred to as Parallax-Interpolation Adaptive Separable Convolution (PIASC),
to generate novel parallax views for the input SSLF in a recursive manner
\cite{gao2018icmew}. 
With regard to the latter type, 
Kalantari \etal propose a learning-based view synthesis method, consisting of disparity and color estimation components, to synthesize novel views for a MLA-based consumer light field camera 
\cite{kalantari2016learning}.
Wu \etal leverage a CNN with a residual learning strategy to perform angular detail restoration on EPIs; however, the maximum disparity range of the input SSLF that can be handled by this method is only 5 pixels  
\cite{wu2017cvpr}. 
More recently, Yeung \etal also exploit an  end-to-end CNN, consisting of the view synthesis and refinement networks, for light field angular resolution enhancement in a coarse-to-fine manner
\cite{yeung2018fast}. 
Nevertheless, this method cannot be directly used to solve the DSLF reconstruction problem because their networks rely on a fixed interpolation rate $\delta$ (see \secref{sec:exp}), while this rate is generally much smaller than the sampling interval $\tau$ (introduced in the next section) for a target DSLF to be reconstructed. 
Wang \etal propose a 4D CNN to enhance the angular resolution of an input 4D SSLF
\cite{wang2018end}; 
however, the interpolation rate $\delta$ of this approach is either 2 or 3 ($\ll\tau$). 
The ST-based IBR algorithm 
\cite{vagharshakyan2018light,vagharshakyan2017accelerated} 
is the first method especially designed for solving the DSLF reconstruction problem. 
In particular, ST fully leverages the light field sparsification in shearlet domain to perform image inpainting on the sparsely-sampled EPIs of the input SSLF
\cite{gao2019mast}. 
Since the proposed learning-based DRST is partially based on the non-learning-based ST, a brief introduction to ST is presented as follows.

\noindent\textbf{Shearlet Transform (ST)} 
\cite{shearlab2016,kutyniok2012shearlets}
is adapted to perform DSLF reconstruction on SSLFs by leveraging the sparsity of EPIs in shearlet domain 
\cite{vagharshakyan2018light,vagharshakyan2017accelerated}. 
Typically, the ST-based DSLF reconstruction comprises four steps: (i) pre-shearing, (ii) shearlet system construction, (iii) sparse regularization and (iv) post-shearing. 
Steps\,(i), (ii) and (iv) require the disparity estimation of the input SSLF, 
\ie the minimal disparity $d_{min}$, maximal disparity $d_{max}$ and disparity range $d_{range}=(d_{max}-d_{min})$.
The estimated disparity data are employed to rearrange the rows of each sparsely-sampled EPI via shearing and zero padding operations and to construct a specifically-tailored universal shearlet system with $\xi$ scales, where $\xi = {\lceil\log_2{\tau\rceil}}$. 
The sparse regularization step is the core of ST, consisting of (i) shearlet analysis transform, (ii) hard thresholding, (iii) shearlet synthesis transform and (iv) double overrelaxation (DORE)
\cite{vagharshakyan2017accelerated}. 
To be more precise, shearlet analysis transform transforms an EPI in image domain into shearlet coefficients in shearlet domain, hard thresholding performs regularization on the transformed coefficients in shearlet domain, shearlet synthesis transform transforms the regularized coefficients into a processed EPI in image domain, 
and DORE is an optional algorithm accelerating the convergence speed of the whole sparse regularization step. 
Moreover, the sparse regularization step is an iterative algorithm, \ie for each color channel of each pre-sheared and zero-padded sparsely-sampled input EPI,
this step is repeated typically 50\,-\,100 times, 
thereby affecting the time efficiency of ST when reconstructing DSLFs from SSLFs of challenging light field scenes that require a high number of iterations.

\begin{figure}[t]
\begin{minipage}[m]{1.\linewidth}
  \centering
  \centerline{\includegraphics[width=1.\textwidth]{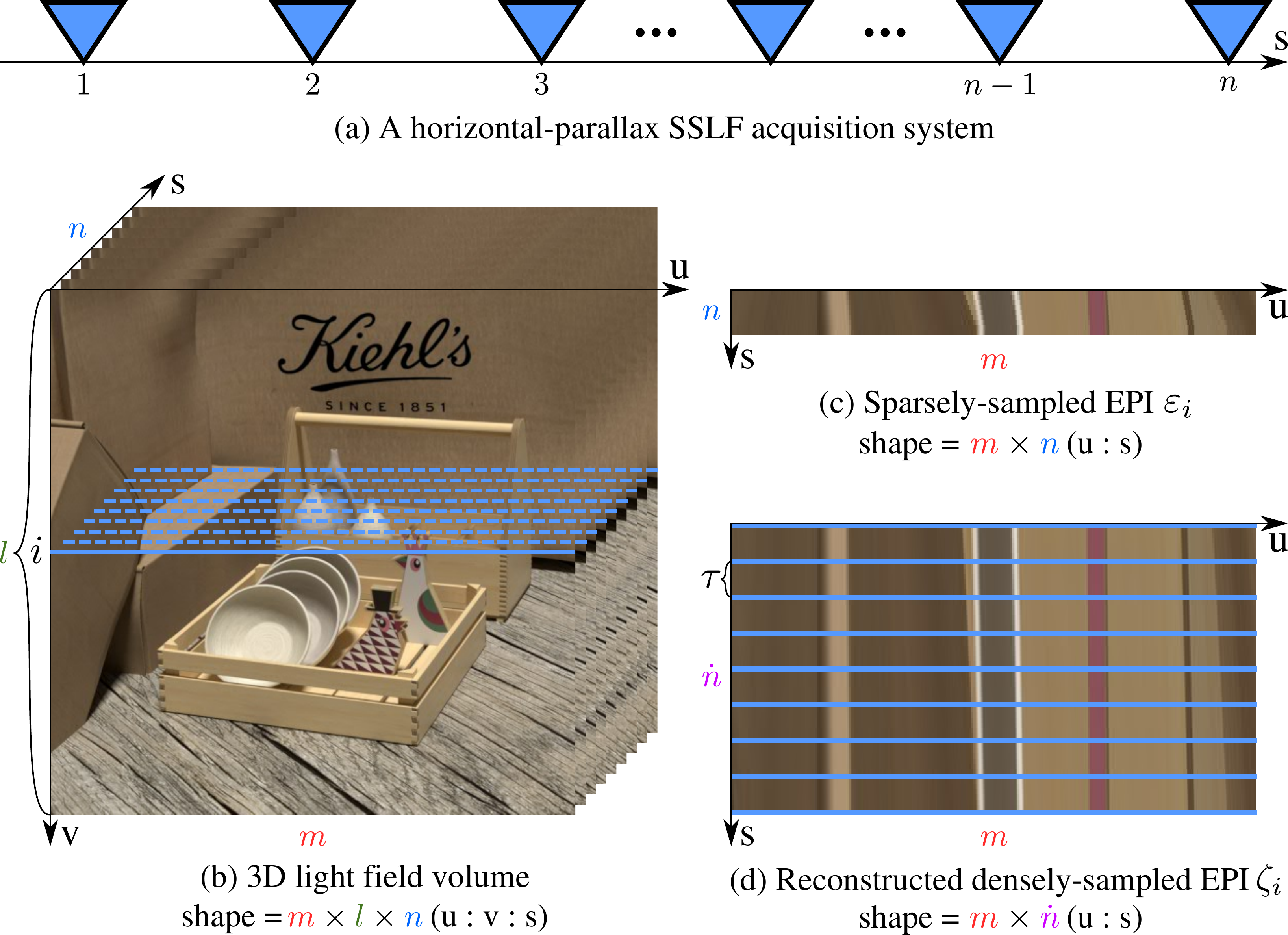}}
\end{minipage}
\vspace{-.8em}
\caption{Introduction to the DSLF reconstruction problem.}
\vspace{.4em}
\label{fig:intro}
\end{figure}

\begin{figure*}[t]
\begin{minipage}[b]{0.33\linewidth}
  \centering
  \centerline{\includegraphics[width=1.\textwidth]{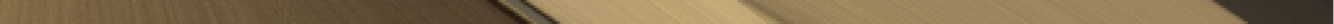}}
%  \vspace{-.1em}
  {\scriptsize(a) Input sparsely-sampled EPI $\varepsilon$ \\ shape = $512 \times 9$ pixels}\medskip
\end{minipage}
\hfill
\begin{minipage}[b]{0.33\linewidth}
  \centering
  \centerline{\includegraphics[width=1.\textwidth]{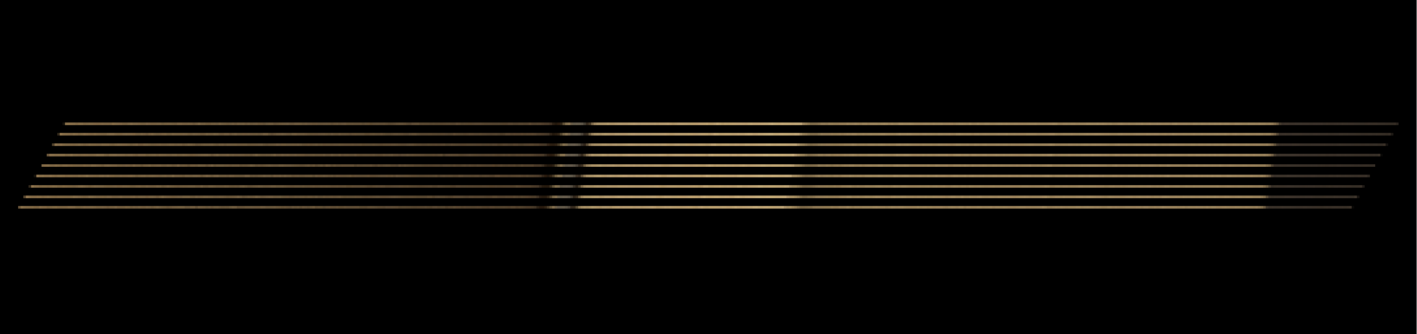}}
%  \vspace{-.1em}
  {\scriptsize(b) Sheared and zero-padded EPI $\dot{\varepsilon}$ \\ shape = $544 \times 128$ pixels}\medskip
\end{minipage}
\hfill
\begin{minipage}[b]{0.291176471\linewidth}
  \centering
  \centerline{\includegraphics[width=1.\textwidth]{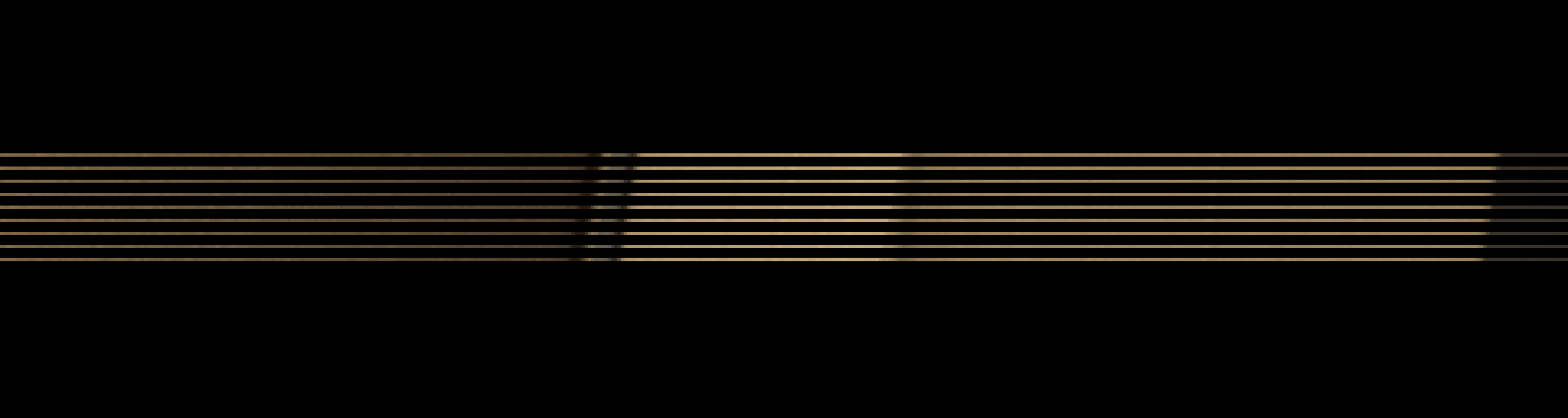}}
%  \vspace{-.1em}
  {\scriptsize(c) Border-cropped EPI $\ddot{\varepsilon}$ \\ shape = $480 \times 128$ pixels}\medskip
\end{minipage}

\vspace{-.3em}

\begin{minipage}[b]{0.24\linewidth}
  \centering
  \centerline{\includegraphics[height=.31372549\textwidth]{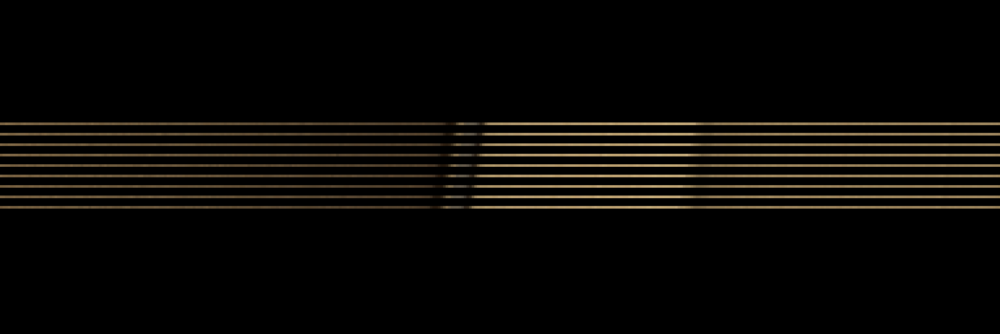}}
%  \vspace{-.1em}
  {\scriptsize(d) Randomly-cropped EPI $\hat{\varepsilon}$\\ shape = $384 \times 128$ pixels}\medskip
\end{minipage}
\hfill
\begin{minipage}[b]{0.24\linewidth}
  \centering
  \centerline{\includegraphics[height=.31372549\textwidth]{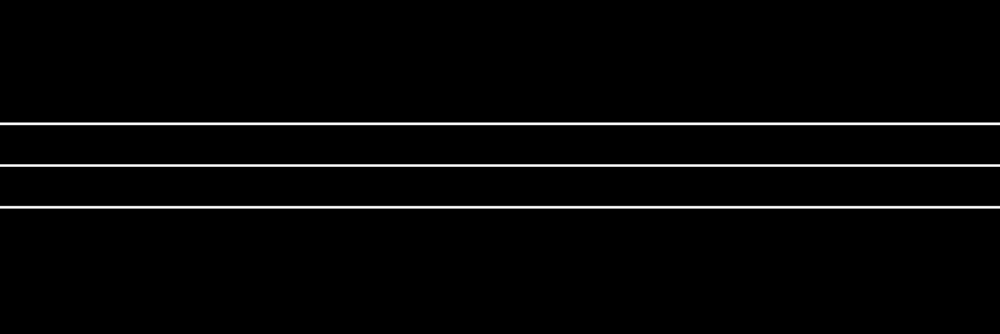}}
%  \vspace{-.1em}
  {\scriptsize(e) Input mask ${\theta}$ \\ shape = $384 \times 128$ pixels}\medskip
\end{minipage}
\hfill
\begin{minipage}[b]{0.24\linewidth}
  \centering
  \centerline{\includegraphics[height=.31372549\textwidth]{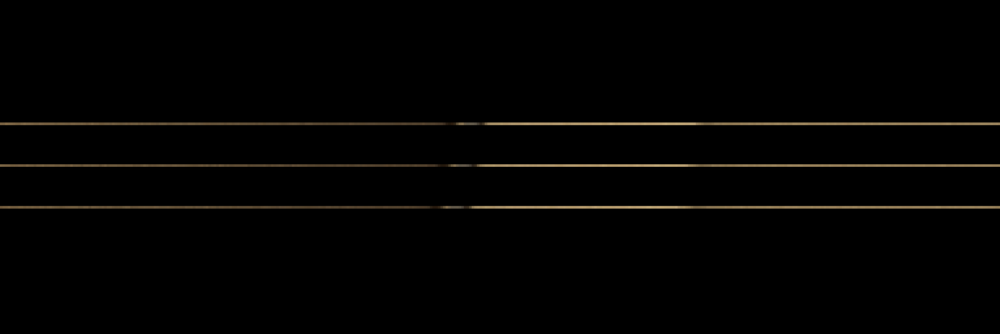}}
%  \vspace{-.1em}
  {\scriptsize(f) $\tau$-decimated EPI $\tilde{\varepsilon}$ ($ = \hat{\varepsilon} \odot \theta$) \\ shape = $384 \times 128$ pixels}\medskip
\end{minipage}
\hfill
\begin{minipage}[b]{0.24\linewidth}
  \centering
  \centerline{\includegraphics[height=.31372549\textwidth]{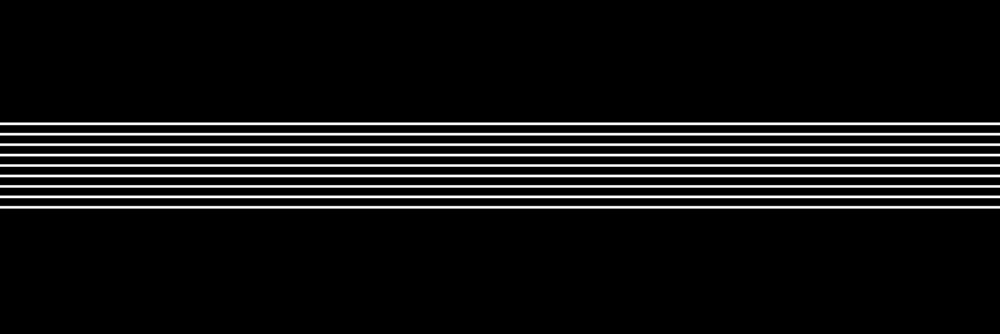}}
%  \vspace{-.1em}
%  {\scriptsize(g) Reconstructed EPI $\tilde{\zeta}$ \\ shape = $384 \times 128$ pixels}\medskip
  {\scriptsize(g) Evaluation mask ${\phi}$ \\ shape = $384 \times 128$ pixels}\medskip
\end{minipage}

\vspace{-1.4em}
\caption{Training data preparation. 
A sparsely-sampled EPI $\varepsilon$ from a training 3D SSLF is illustrated in (a). 
The sheared and zero-padded EPI $\dot{\varepsilon}$ in (b) is the result of performing the pre-shearing and zero-padding step on $\varepsilon$. 
The border of $\dot{\varepsilon}$ is cut to generate a border-cropped EPI $\ddot{\varepsilon}$ in (c).
A random cropping operation is then performed on $\ddot{\varepsilon}$ to produce a $3:1$ randomly-cropped EPI $\hat{\varepsilon}$ presented in (d).
For $\hat{\varepsilon}$, an input mask $\theta$ is designed as shown in (e) and utilized to generate the $\tau$-decimated EPI $\tilde{\varepsilon}$ in (f), which is the input data for the learning-based sparse regularization step of DRST.
Finally, the evaluation mask $\phi$ in (g) is employed to calculate the loss function \eqref{eq:loss} of the learning-based sparse regularization step.
}
\label{fig:train}
\vspace{.4em}
\end{figure*}

\vspace{-.6em}
\section{Deep Residual Shearlet Transform (DRST)}	\label{sec:method}

%these two methods are introduced in this section, respectively.

Inspired by the above ST-based DSLF reconstruction, a novel learning-based ST approach, referred to as DRST, is proposed by fully leveraging the state-of-the-art deep learning techniques. 
Specifically, DRST also consists of four steps: 
(i) shearlet system construction, 
(ii) pre-shearing and zero-padding, 
(iii) learning-based sparse regularization and 
(iv) post-shearing.
The details of these four steps will be elaborated after defining the light field-associated symbols and notations.

\subsection{Symbols and notations}
As illustrated in \figref{fig:intro}\,(a), an input horizontal-parallax SSLF is essentially a set of images uniformly sampled along the horizontal axis s. 
After stacking all the images of the input 3D SSLF along axis s, a 3D light field volume can be generated  as shown in \figref{fig:intro}\,(b). 
The generated 3D light field volume has a spatial resolution of $m\times l$ pixels and an angular resolution of $n$ pixels.
Let the input 3D SSLF be denoted by $\mathcal{S}=\{\varepsilon_i | 1 \leqslant i \leqslant l\}$, where $\varepsilon_i \in \mathbb{R}^{m \times n \times 3}$ represents a sparsely-sampled EPI. 
To better understand the EPI structure, one of the sparsely-sampled EPIs of $\mathcal{S}$, \ie $\varepsilon_i$, is picked up from the 3D light field volume in \figref{fig:intro}\,(b) and shown in \figref{fig:intro}\,(c).
Similarly, the target DSLF to be reconstructed from $\mathcal{S}$ is represented by $\mathcal{D}=\{\zeta_i | 1 \leqslant i \leqslant l\}$, where $\zeta_i \in \mathbb{R}^{m \times \dot{n} \times 3}$ stands for a densely-sampled EPI.
It should be noted that each densely-sampled EPI in $\mathcal{D}$ is reconstructed from a corresponding sparsely-sampled EPI in $\mathcal{S}$.
The densely-sampled EPI $\zeta_i$, corresponding to $\varepsilon_i$, is presented in \figref{fig:intro}\,(d).
Comparing these two EPIs, it can be found that $\zeta_i$ has a higher resolution than $\varepsilon_i$ along the s axis. 
Specifically, the number of rows of $\zeta_i$, \ie $\dot{n}$, is decided by the sampling interval $\tau$ and the number of rows of $\varepsilon_i$, \ie $n$, with an equation $\dot{n}=\big((n-1)\tau+1\big)$. 
In other words, for the same input SSLF $\mathcal{S}$, the angular resolution of the target DSLF $\mathcal{D}$ to be reconstructed depends on the sampling interval $\tau$ that is controlled by the disparity range ($d_{range}$) of $\mathcal{S}$, \ie $\tau \geqslant d_{range}$.
In this paper, we target solving the DSLF reconstruction problem for any input SSLF $\mathcal{S}$ with a moderate disparity range, \ie $8 < d_{range} \leqslant 16$ pixels.
In addition, as mentioned in the previous section, the number of the scales of the target shearlet system, $\xi$, relies on the sampling interval $\tau$.
In particular, $\xi = 4$ when $ 8 < \tau \leqslant 16$, and $\xi = 5$ when $ 16 < \tau \leqslant 32$.  
For the shearlet system construction and learning-based sparse regularization steps of DRST, 
using a shearlet system with 4 scales is much faster than using a shearlet system with 5 scales. 
As a result, the sampling interval $\tau$ is set to $16$ for this paper.

%\vspace{-.9em}
\subsection{Shearlet system construction} 
%\vspace{-.25em}
The specifically-tailored universal shearlet system in
\cite{vagharshakyan2018light} is chosen to be constructed for the shearlet analysis and synthesis transforms in the learning-based sparse regularization step.
A shearlet analysis transform is denoted by $\mathcal{SH} : \mathbb{R}^{\gamma \times \gamma} \rightarrow \mathbb{R}^{\gamma \times \gamma \times \eta}$, where $\gamma \times \gamma$ represents the size of a shearlet filter and $\eta$ denotes the number of shearlets in a shearlet system. 
A shearlet synthesis transform is represented by $\mathcal{SH}^* : \mathbb{R}^{\gamma \times \gamma \times \eta} \rightarrow \mathbb{R}^{\gamma \times \gamma}$.
Note that the number of shearlets, \ie $\eta$, is decided by the number of the scales, \ie $\xi$, of the target shearlet system with an equation $\eta = (2^{\xi+1}+\xi-1)$. 
In addition, as described in the previous section, $\xi$ is decided by the sampling interval $\tau$.
In our case, $\xi={\lceil\log_2{\tau\rceil}}=4$ and,
consequently, the target shearlet system has $\eta=35$ shearlets.  
The size of the shearlet filters in the target shearlet system is specified by the users, \ie $\gamma = 127$ for this paper.
% For further details, refer to 
%\cite{genzel2014asymptotic}. 

\begin{figure*}[t]
\hfill
\begin{minipage}[b]{1.\linewidth}
  \centering
  \centerline{\includegraphics[width=.86\textwidth]{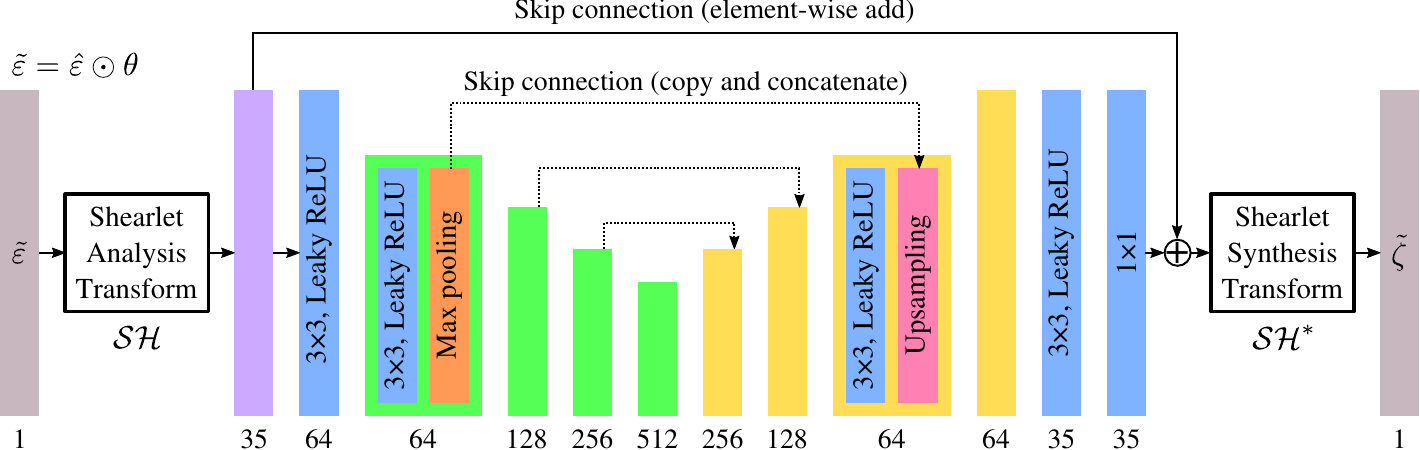}}
\end{minipage}
\hfill
\vspace{-1.4em}
\caption{Network architecture of the learning-based sparse regularization in DRST.}
\label{fig:architecture}
\vspace{.4em}
\end{figure*}

%\vspace{-1.2em}
\subsection{Pre-shearing and zero-padding} 
\label{sec:pre-sh-pad}
For better understanding the pre-shearing and zero-padding strategies and how to leverage the synthetic SSLF data for training,
in this section the first-row horizontal-parallax light field of the 4D light field ``Boxes'' %in our training dataset (see \secref{sec:exp})
%the 4D light field dataset 
\cite{honauer2016dataset} 
is selected as the input 3D SSLF $\mathcal{S}$ for demonstration. 
The input 3D light field $\mathcal{S}$ has an angular resolution 9 pixels and a spatial resolution $512\times 512$ pixels.
The ground-truth disparity information of $\mathcal{S}$ is provided by the dataset, \ie $d_{min}=-2.2$, $d_{max}=1.4$ and $d_{range}=3.6$ pixels.
The first sparsely-sampled EPI of $\mathcal{S}$, represented by $\varepsilon$, is illustrated in \figref{fig:train}\,(a).
It can be seen that the shape of $\varepsilon$ is $512\times9$ pixels.
The values of $d_{min}$ and $d_{range}$ are utilized to shear and pad $\varepsilon$ as shown in \figref{fig:train}\,(b). 
Specifically, the sheared and zero-padded EPI $\dot{\varepsilon}$ has nine separated non-black lines from $\varepsilon$. 
The horizontal and vertical displacements between neighboring non-black lines are $\varphi$ and $\frac{\tau}{4}=4$ pixels, %$\lceil d_{range}\rceil$ ($=4$), 
respectively.
Here, $ \big(d_{min}-(\frac{\tau}{4}-d_{range})\big) \leqslant \varphi \leqslant d_{min}$ and $d_{range} \leqslant \frac{\tau}{4}$, 
are such that the image inpainting on $\dot{\varepsilon}$ can produce a densely-sampled EPI.
%Note that the $d_{range}$ of the training SSLF should be less than or equal to the $\frac{\tau}{4}$\,-\,pixels vertical displacement. 
Moreover, the size of $\dot{\varepsilon}$ is $544\times 128$ pixels.
The left and right borders of $\dot{\varepsilon}$ are then cut to generate a border-cropped EPI $\ddot{\varepsilon}$ shown in \figref{fig:train}\,(c) with a shape $480 \times 128$ pixels.
In order to augment training data, a $384\times 128$\,-\,pixels EPI $\hat{\varepsilon}$ is randomly cropped from $\ddot{\varepsilon}$ for each training iteration.
Note that $\ddot{\varepsilon}$ and $\hat{\varepsilon}$ have the same height, implying that the random cropping operation here is essentially to slide a 3\,:\,1 window inside $\ddot{\varepsilon}$ along the horizontal axis to produce one crop, \ie $\hat{\varepsilon}$, which is illustrated in \figref{fig:train}\,(d). 	

\noindent\textbf{Masks.}
There are two masks, input mask $\theta$ and evaluation mask $\phi$, associated with the cropped EPI $\hat{\varepsilon}$ and prepared for the learning-based sparse regularization step. 
In particular, the input mask $\theta$ has three non-zero lines, corresponding to the first, middle and last non-zero lines of $\hat{\varepsilon}$.
The evaluation mask $\phi$ has nine non-zero lines, corresponding to all the nine non-zero lines of $\hat{\varepsilon}$.
The input mask $\theta$ and evaluation mask $\phi$ are illustrated in \figref{fig:train}\,(e) and (g), respectively.
In addition,
the input mask $\theta$ is utilized to generate a sparsely-sampled EPI $\tilde{\varepsilon}$ from the cropped EPI $\hat{\varepsilon}$ as the input data for the sparse regularization of ST and learning-based sparse regularization of DRST, \ie $\tilde{\varepsilon} = \hat{\varepsilon} \odot \theta$, 
where $\odot$ denotes the Hadamard product.
The vertical displacement between any two adjacent non-zero lines of the input mask $\theta$ or sparsely-sampled $\tilde{\varepsilon}$ is equal to the sampling interval $\tau$.
As a result, $\tilde{\varepsilon}$ is also referred to as $\tau$-decimated EPI, which is illustrated in \figref{fig:train}\,(f). 

\vspace{-1.2em}
\subsection{Learning-based sparse regularization}
\label{sec:sparsity_regu}
The goal of the sparse regularization step in ST is to reconstruct a densely-sampled EPI $\tilde{\zeta}$ from the above generated $\tau$-decimated EPI $\tilde{\varepsilon}$. This can be achieved by solving the following optimization problem in the shearlet transform domain:
\begin{IEEEeqnarray}{c}
\underset{\tilde{\zeta}}{\min}~\norm{\mathcal{SH}(\tilde{\zeta})}_1,~\text{s.t.}~\tilde{\varepsilon}=\theta\odot\tilde{\zeta}\,.
\end{IEEEeqnarray}
The sparse regularization is an iterative algorithm that solves the above problem by performing regularization on the transform domain coefficients. 
Different from the sparse regularization step of ST, the learning-based sparse regularization step of DRST is a more efficient non-iterative algorithm, 
which is introduced as below:

\noindent\textbf{Network architecture.}
The learning-based sparse regularization in DRST is a deep CNN consisting of an encoder-decoder network and a residual learning strategy, which are inspired by U-Net 
\cite{ronneberger2015u} and ResNet
\cite{he2016deep}, respectively.
The network architecture of this CNN is presented in \figref{fig:architecture}. 
As shown in this figure, the input data is the $\tau$-decimated EPI $\tilde{\varepsilon}$ and the output data is the reconstructed densely-sampled EPI $\tilde{\zeta}$. 
The shearlet analysis transform converts $\tilde{\varepsilon}$ into $35$\,-\,channels shearlet coefficients in shearlet domain.
These coefficients are then fed to the encoder-decoder network to predict residual shearlet coefficients.  
Specifically, the encoder-decoder network is a U-Net with an encoder having 4 hierarchies and a decoder also having 4 hierarchies. 
The encoder and decoder in the U-Net are connected by three skip connections (copy and concatenate) at the same spatial resolution for the first three hierarchies. 
Each hierarchy in the encoder is composed of three layers, \ie a 2D convolution layer, a Leaky ReLU layer ($\alpha = 0.3$) and a max pooling layer for decreasing the spatial resolution by 2.
Each hierarchy in the decoder also consists of three layers, \ie a 2D convolution layer, a Leaky ReLU layer ($\alpha = 0.3$) and an upsampling layer with nearest interpolation for increasing the spatial resolution by 2. 
The convolution kernel size is set to $3\times 3$ for all the 2D convolution layers except for the last one, where the convolution kernel size is set to $1\times 1$. In addition, no Leaky ReLU layer is added behind the last 2D convolution layer.
Afterwards,
the residual learning strategy is utilized to add the predicted residual shearlet coefficients back to the original shealet coefficients by means of the other type of skip connection, \ie an element-wise add operation.
Finally, these processed shearlet coefficients are transformed back to image domain to generate $\tilde{\zeta}$ via the shearlet synthesis transform. 
Mathematically, the learning-based sparse regularization can be written as below:
\begin{IEEEeqnarray}{c}
\tilde{\zeta} = \mathcal{SH}^*\Big(\mathcal{SH}(\tilde{\varepsilon}) + \mathcal{R}\big(\mathcal{SH}(\tilde{\varepsilon})\big) \Big)\,,
\end{IEEEeqnarray}
where $\mathcal{R}$ denotes the encoder-decoder network, 
\ie $\mathcal{R} : \mathbb{R}^{128 \times 384 \times 35} \rightarrow \mathbb{R}^{128 \times 384 \times 35}$ for the training case.

\begin{figure*}[t]

\begin{minipage}[t]{0.195\linewidth}
  \centering
  \centerline{\includegraphics[width=1.\textwidth]{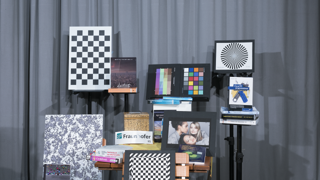}}
  \vspace{-.4em}
  \centerline{\scriptsize(a) $\Psi_1^1$: Books and charts}\medskip
\end{minipage}
\hfill
\begin{minipage}[t]{0.195\linewidth}
  \centering
  \centerline{\includegraphics[width=1.\textwidth]{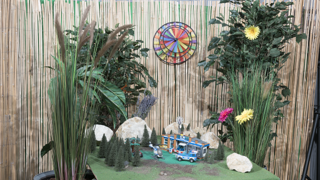}}
  \vspace{-.4em}
  \centerline{\scriptsize(b) $\Psi_2^1$: Lego city}\medskip
\end{minipage}
\hfill
\begin{minipage}[t]{0.195\linewidth}
  \centering
  \centerline{\includegraphics[width=1.\textwidth]{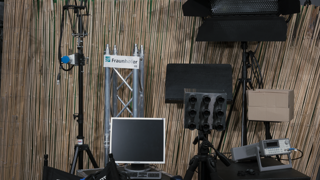}}
  \vspace{-.4em}
  \centerline{\scriptsize(c) $\Psi_3^1$: Lightfield production}\medskip
\end{minipage}
\hfill
\begin{minipage}[t]{0.195\linewidth}
  \centering
  \centerline{\includegraphics[width=1.\textwidth]{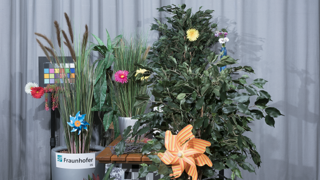}}
  \vspace{-.4em}
  \centerline{\scriptsize(d) $\Psi_4^1$: Plants}\medskip
\end{minipage}
\hfill
\begin{minipage}[t]{0.195\linewidth}
  \centering
  \centerline{\includegraphics[width=1.\textwidth]{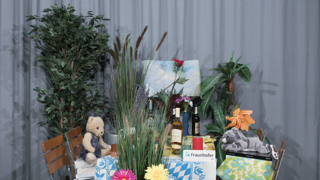}}
  \vspace{-.4em}
  \centerline{\scriptsize(e) $\Psi_5^1$: Table in the garden}\medskip
\end{minipage}

\vspace{-.3em}

\begin{minipage}[t]{0.195\linewidth}
  \centering
  \centerline{\includegraphics[width=1.\textwidth]{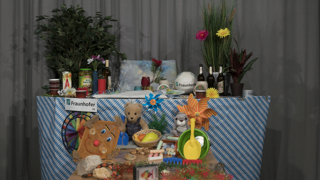}}
  \vspace{-.4em}
  \centerline{\scriptsize(f) $\Psi_6^1$: Table top \RNum{1}}\medskip
\end{minipage}
\hfill
\begin{minipage}[t]{0.195\linewidth}
  \centering
  \centerline{\includegraphics[width=1.\textwidth]{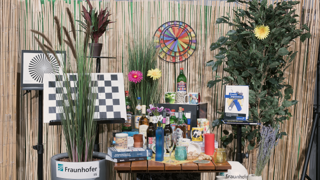}}
  \vspace{-.4em}
  \centerline{\scriptsize(g) $\Psi_7^1$: Table top \RNum{2}}\medskip
\end{minipage}
\hfill
\begin{minipage}[t]{0.195\linewidth}
  \centering
  \centerline{\includegraphics[width=1.\textwidth]{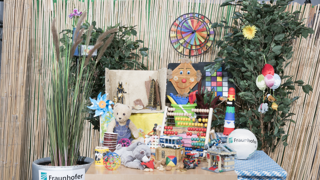}}
  \vspace{-.4em}
  \centerline{\scriptsize(h) $\Psi_8^1$: Table top \RNum{3}}\medskip
\end{minipage}
\hfill
\begin{minipage}[t]{0.195\linewidth}
  \centering
  \centerline{\includegraphics[width=1.\textwidth]{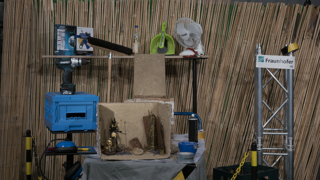}}
  \vspace{-.4em}
  \centerline{\scriptsize(i) $\Psi_9^1$: Workshop}\medskip
\end{minipage}
\hfill
\begin{minipage}[t]{0.195\linewidth}
  \centering
  \centerline{\includegraphics[height=0.5625\textwidth]{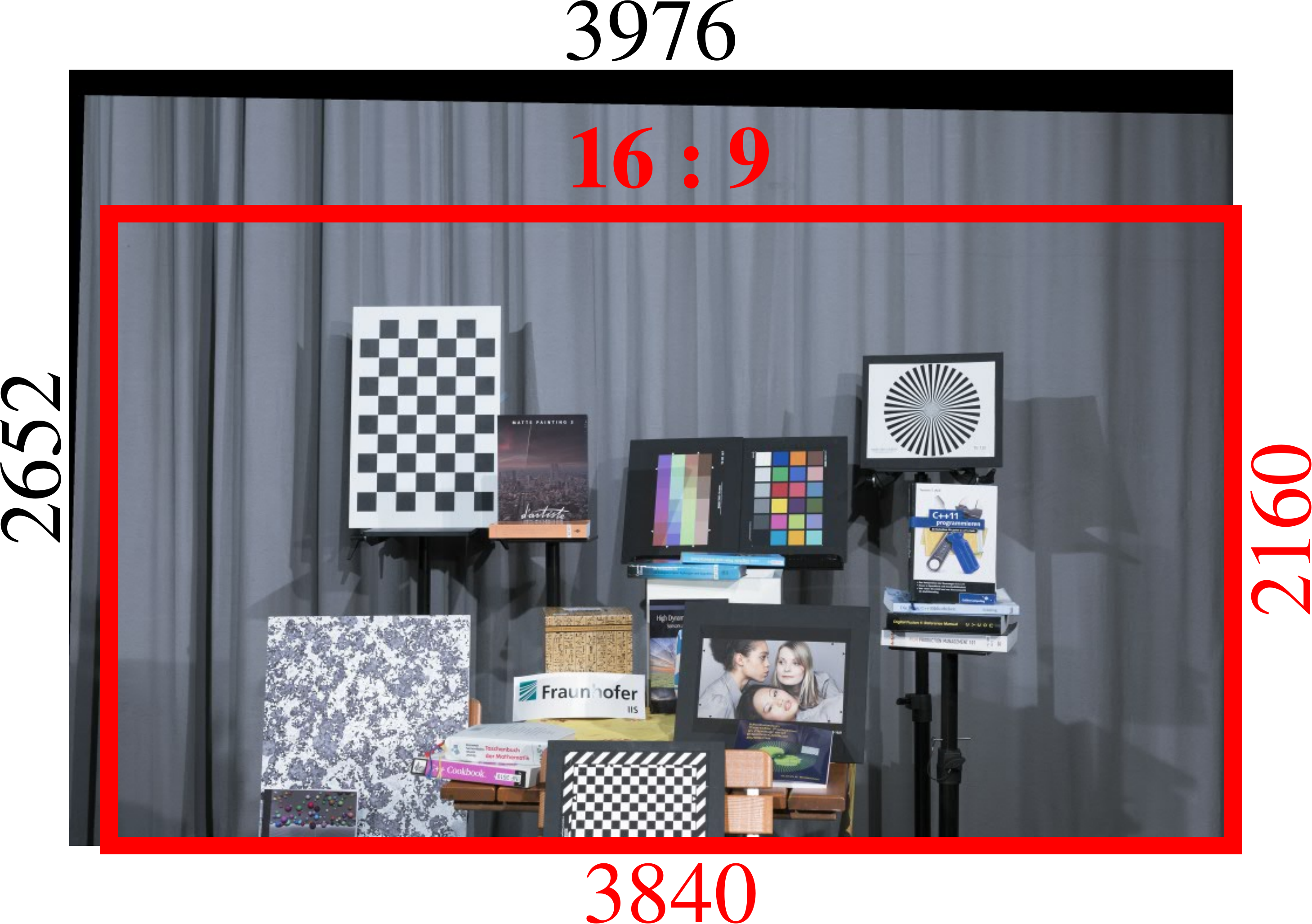}}
  \vspace{-.4em}
  \centerline{\scriptsize(j) Cutting and scaling strategy}\medskip
\end{minipage}

\vspace{-.8em}
\caption{Middle views of ground-truth light fields ${\Psi}_e^1$ ($1 \leqslant e \leqslant 9$) in the evaluation dataset 1. Sub-image (j) illustrates the image cutting and scaling strategy in \secref{sec:eva_data_1}. }
\label{fig:dataset}
%\vspace{.4em}
\end{figure*}

\begin{table*}[!t]
\begin{center}
\caption{Disparity estimation, minimum and average per-view PSNR results (in dB, explained in \secref{sec:qua}) for the performance evaluation of different light field reconstruction methods on the evaluation dataset 1. %evaluation datasets $\mathcal{S}_e$ ($1 \leqslant e \leqslant 9$). 
} 
\label{tab:res1}
\vspace{-.8em}
\scalebox{1.}{
\begin{tabular}{c||c|c|c|c|c|c|c}
\specialrule{.15em}{.0em}{.2em}
\multirow{2}{*}{$e$ of $\mathcal{S}_e^1$} & \multicolumn{3}{c|}{\textbf{Disparity Estimation (pixel)} } & \multicolumn{4}{c}
{\textbf{Minimum PSNR / Average PSNR (dB)} } \\ [.5ex] 
\hhline{|~||-------} 
  & $d_{min}^\mathcal{S}$ & $d_{max}^\mathcal{S}$ & $d_{range}^\mathcal{S}$  & SepConv ($\mathcal{L}_1$) \cite{niklaus2017iccv} & PIASC ($\mathcal{L}_1$) \cite{gao2018icmew} & ST \cite{vagharshakyan2017accelerated} & DRST 
  \\ [.1ex] 
  \hline
1 & 12.5 & 22 & 9.5   & 21.963 / 24.907 & 21.957 / 24.905 & 35.277 / 38.611 & \textbf{38.241} / \textbf{39.933} \\  
2 & 13.5 & 24.5 & 11  & 26.562 / 29.073 & 26.579 / 29.117 & 27.376 / \textbf{29.831} & \textbf{27.698} / 29.820 \\
3 & 14 & 27.5 & 13.5  & 30.508 / 32.874 & 30.528 / 32.952 & \textbf{32.092} / \textbf{34.221} & 31.568 / 33.133 \\
4 & 12.5 & 27.5 & 15  & 31.536 / 34.804 & 31.584 / 34.986 & 32.603 / \textbf{36.258} & \textbf{33.519} / 36.220 \\
5 & 12.5 & 27 & 14.5  & 32.278 / 33.803 & 32.327 / 33.926 & 32.812 / \textbf{35.372} & \textbf{34.020} / 35.239 \\
6 & 12.5 & 27 & 14.5  & 30.100 / 32.539 & 30.108 / 32.605 & 36.412 / 40.423 & \textbf{40.237} / \textbf{41.505} \\
7 & 13 & 21.5 & 8.5   & 26.609 / 29.939 & 26.621 / 29.982 & 28.126 / 30.367 & \textbf{28.973} / \textbf{30.759} \\
8 & 14 & 24.5 & 10.5  & 26.885 / 29.480 & 26.910 / 29.536 & 28.165 / 30.312 & \textbf{29.529} / \textbf{31.061} \\ 
9 & 14 & 27.5 & 13.5  & 33.043 / 35.533 & 33.078 / 35.672 & 34.242 / \textbf{36.619} & \textbf{35.004} / 36.177 \\
%\hline 
\specialrule{.15em}{.2em}{.0em}
\end{tabular}
}
\end{center}
%\vspace{-.8em}
\end{table*}

\noindent\textbf{Loss function.} 
The trainable parameters in $\mathcal{R}$ are learned by solving the following optimization problem:
\begin{IEEEeqnarray}{c}
\underset{\mathcal{R}}{\min}~\norm{ \tilde{\zeta} - \tilde{\zeta}^\text{GT}}_1\,.%~\text{s.t.}~\tilde{\varepsilon}=\theta\odot\tilde{\zeta}\,.
\end{IEEEeqnarray}
However, the ground-truth densely-sampled EPI $\tilde{\zeta}^\text{GT}$ corresponding to the reconstructed densely-sampled EPI $\tilde{\zeta}$ is unknown, 
since the training synthetic SSLF dataset does not offer the corresponding ground-truth DSLF data.
Besides, rendering a high-quality and high-resolution synthetic DSLF dataset is  prohibitively expensive compared to the rendering of a synthetic SSLF dataset.
%real-world DSLFs are hard to capture as is pointed out in the introduction to this paper.
Therefore, a new loss function without relying on the DSLF data is proposed. % and presented as following:
Specifically, the loss function for the training of the encoder-decoder network in the learning-based sparse regularization takes account of minimizing the reconstruction error between the ground-truth sparsely-sampled EPI $\hat{\varepsilon}$ and the reconstructed densely-sampled EPI $\tilde{\zeta}$ using the evaluation mask $\phi$ via $\ell_1$ norm, \ie
\begin{IEEEeqnarray}{c}	\label{eq:loss}
\mathcal{L} =  \norm{\hat{\varepsilon} - \tilde{\zeta} \odot \phi}_1 \,.
\end{IEEEeqnarray}
Although the ground-truth sparsely-sampled EPI $\hat{\varepsilon}$ is not densely-sampled, 
it contains 6 non-zero lines that the input $\tau$-decimated EPI $\tilde{\varepsilon}$ does not have, thereby guiding the optimization process for the training of the network of the learning-based sparse regularization.

%where 
%\begin{IEEEeqnarray}{c}
%\mathcal{L}^\texttt{SRN} =  \norm{(\varepsilon'_i - \zeta'_i)\odot(\theta'_i - \theta_i)}_1 \,, \IEEEyesnumber\IEEEyessubnumber*\\
%\mathcal{L}^\texttt{CCN} =  \norm{(\varepsilon'_i - \zeta''_i)\odot\theta'_i}_1 \,.
%%\vspace{-.4em}
%\end{IEEEeqnarray}
%\vspace{-1.2em}
\subsection{Post-shearing} 
%The reconstructed densely-sampled EPI $\tilde{\zeta}$ corresponding to the source sparsely-sampled EPI $\tilde{\varepsilon}$ is illustrated in \figref{fig:train}\,(g).
The target DSLF can be reconstructed after compensating for the horizontal displacement produced by the aforementioned pre-shearing strategy, \ie ${\varphi}$ described in \secref{sec:pre-sh-pad}, for all the reconstructed densely-sampled EPIs. 
More details can also be found in \cite{gao2019light}.

%\vspace{-.6em}
\section{Experiments}	\label{sec:experiment}
\subsection{Experimental Settings} 	\label{sec:exp}
As explained in the introduction section,
the proposed DRST approach is trained on a synthetic SSLF dataset with ground-truth disparity information and evaluated on three challenging real-world light field evaluation datasets. 
Since both ST and DRST are designed for reconstructing light fields that are densely-sampled, 
there are two requirements, 
\ie (a) $d_{range}\leqslant\frac{\tau}{4}$ and (b) $d_{range}^{\mathcal{S}}\leqslant \tau$,
for the training and evaluation datasets, respectively.
Besides, 
the interpolation rate $\delta$ denotes the sampling rate for extracting a SSLF $\mathcal{S}$ from a ground-truth light field $\Psi$ in an evaluation dataset
\cite{gao2019iest}. 
More details about the preparation of the training and evaluation datasets and the implementation of DRST are presented next.
%The reason for this setup is that a synthetic SSLF dataset can provide ground-truth disparity information, which is convenient for the pre-shearing step of the network training of DRST. 
%Besides, the reason for choosing three real-world SSLF datasets for evaluation is that these datasets are much more challenging than the synthetic one.    

\subsubsection{Training dataset} \label{sec:train_data}
The 4D light field dataset 
\cite{honauer2016dataset} 
is a synthetic dataset created with Blender. It is composed of 28 4D light fields of the same size, \ie $9\times9\times512\times512\times3$. 
Among them, there are 18 4D light fields suitable for the network training of DRST, since 
(i) the four light fields in the category ``Stratified'' differ a lot from real-world light fields; 
%Among the remaining $24$ 4D light fields, 
%the ``Museum'' 4D light field is also discarded, 
(ii) the 4D light field ``Museum'' is rendered for a non-Lambertian scene, where the shadows on the glass lead to a real $d_{min}$ that is lower than the ground-truth $d_{min}$ provided by the dataset;
% make the real $d_{range}$ much larger than the ground-truth $d_{range}$ provided by the dataset.
(iii) the 4D light fields ``Herbs'', ``Antinous'', ``Dishes'', ``Greek'' and ``Tower'' do not satisfy the requirement (a).
%that $d_{range}\leqslant\frac{\tau}{4}$ in the pre-shearing and zero-padding step described in \secref{sec:pre-sh-pad}.
The 18 suitable 4D light fields are split into both horizontal- and vertical-parallax SSLFs for a total of
$18\times (9+9)=324$ sets. 
Note that all the vertical-parallax SSLFs are turned into horizontal-parallax SSLFs by performing $90\degree$ anticlockwise rotation on all the parallax images. 
The generated 3D SSLFs $\mathcal{S}_t$ ($1 \leqslant t \leqslant 324$) have the same angular and spatial resolutions, \ie $n=9$, $l=512$ and $m=512$.  
To augment the number of training samples, the pre-shearing strategy in \secref{sec:pre-sh-pad} is  repeated three times for each $\mathcal{S}_t$, 
corresponding to the horizontal displacements $\varphi = d_{min}$, $d_{min}-0.5\cdot(\frac{\tau}{4}-d_{range})$ and $d_{min}-(\frac{\tau}{4}-d_{range})$, respectively.
As a result, 972 sheared input SSLFs are generated, producing $972\times 512 = 497,664$ border-cropped EPIs for training, of which an example is displayed in \figref{fig:train}\,(c) .

%The sampling interval $\tau^t = \lceil d_{range}^t\rceil$ (see \secref{sec:intro}) depends on the ground-truth $d_{range}^t$ offered by the dataset. It varies from $3$ to $8$ pixels across different training SSLFs. 
%As a result, the target DSLF $\mathcal{D}_t$ to be reconstructed from $\mathcal{S}_t$ has $\dot{n}^t=(8\tau^t+1)$ horizontal-parallax views.
%It should be noted that, due to the usage the elaborately-designed input logical mask $\theta_i$ as shown in \figref{fig:pipeline}\,(b), the real sampling interval $\tilde{\tau}^t$ of the training data for DRST training is four times as large as $\tau^t$, which means that  the real sampling interval $\tilde{\tau}^t$ varies from $12$ to $32$ pixels. 
%In \secref{sec:dcst}, it is mentioned that the constructed shearlet system is designed for dealing with DSLF reconstruction \textit{w.r.t.} the case of $\tilde{\tau}^t \leqslant 16$ pixels. 
%However, we find that feeding training data that are beyond the capability of the constructed shearlet system to the network of DRST can accelerate the network parameter learning of DRST. 
%%increasing the difficulty of the training data here can help the network training of DRST. 
%Besides, $\tilde{\tau}^t$ is still much smaller than the spacial resolution of the constructed shearlet system, \ie $\tilde{\tau}^t \ll \gamma$.
%%which will be introduced in implementation details part. 
%An example of a training SSLF $\mathcal{S}_t$ is displayed in \figref{fig:intro}\,(b). 

\begin{figure*}[t]

\begin{minipage}[t]{0.195\linewidth}
  \centering
  \centerline{\includegraphics[width=1.\textwidth]{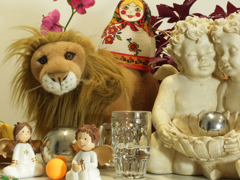}}
  \vspace{-.4em}
  \centerline{\scriptsize(a) $\Psi_1^2$: Toys}\medskip
\end{minipage}
\hfill
\begin{minipage}[t]{0.195\linewidth}
  \centering
  \centerline{\includegraphics[width=1.\textwidth]{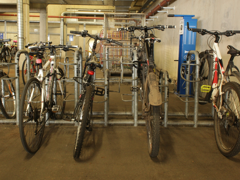}}
  \vspace{-.4em}
  \centerline{\scriptsize(b) $\Psi_2^2$: Bikes}\medskip
\end{minipage}
\hfill
\begin{minipage}[t]{0.195\linewidth}
  \centering
  \centerline{\includegraphics[width=1.\textwidth]{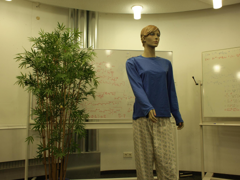}}
  \vspace{-.4em}
  \centerline{\scriptsize(c) $\Psi_3^2$: Mannequin}\medskip
\end{minipage}
\hfill
\begin{minipage}[t]{0.195\linewidth}
  \centering
  \centerline{\includegraphics[width=1.\textwidth]{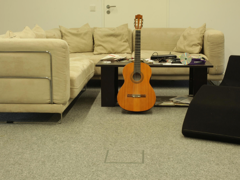}}
  \vspace{-.4em}
  \centerline{\scriptsize(d) $\Psi_4^2$: Livingroom}\medskip
\end{minipage}
\hfill
\begin{minipage}[t]{0.195\linewidth}
  \centering
  \centerline{\includegraphics[width=1.\textwidth]{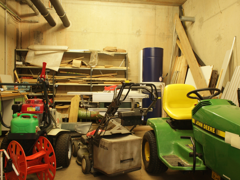}}
  \vspace{-.4em}
  \centerline{\scriptsize(e) $\Psi_5^2$: Workshop}\medskip
\end{minipage}

\vspace{-.8em}
\caption{Middle views of ground-truth light fields ${\Psi}_e^2$ ($1 \leqslant e \leqslant 5$) in the evaluation dataset 2.}
%\vspace{.4em}
\label{fig:dataset2}
\end{figure*}

\begin{table*}[!t]
\begin{center}
\caption{Disparity estimation, interpolation rate, minimum and average per-view PSNR results (in dB, explained in \secref{sec:qua}) for the performance evaluation of different light field reconstruction methods on the evaluation dataset 2. %evaluation datasets $\mathcal{S}_e$ ($1 \leqslant e \leqslant 9$). 
} 
\label{tab:res2}
\vspace{-.8em}
\scalebox{1.}{
\begin{tabular}{c||c|c|c|c|c|c|c|c}
\specialrule{.15em}{.0em}{.2em}
\multirow{2}{*}{$e$ of $\Psi_e^2$} & \multicolumn{3}{c|}{\textbf{Disparity Estimation (pixel)} } & \textbf{Interpolation rate} & \multicolumn{4}{c}
{\textbf{Minimum PSNR / Average PSNR (dB)} } \\ [.5ex] 
\hhline{~||---~|----} 
  & $d_{min}^\Psi$ & $d_{max}^\Psi$ & $d_{range}^\Psi$ & $\delta^2$ & SepConv ($\mathcal{L}_1$) \cite{niklaus2017iccv} & PIASC ($\mathcal{L}_1$) \cite{gao2018icmew} & ST \cite{vagharshakyan2017accelerated} & DRST 
  \\ [.1ex] 
  \hline
1 & -1.03125 & 1.0625 & 2.09375  & 4 & 36.427 / 41.554 & \textbf{36.500} / \textbf{41.769} & 36.215 / 40.233 & 36.492 / 41.045 \\
2 & -0.875 & 0.59375 & 1.46875   & 8 & 33.765 / 36.597 & 33.884 / \textbf{36.665} & 32.876 / 36.092 & \textbf{33.971} / 36.435 \\
3 & -0.46875 & 0.4375 & 0.90625  & 16 & 35.757 / 37.658 & 35.795 / \textbf{37.915} & 34.426 / 37.433 & \textbf{35.814} / 37.847 \\
4 & -0.375 & 0.5 & 0.875         & 16 & 40.507 / 42.867 & \textbf{40.636} / \textbf{43.631} & 39.064 / 42.028 & 40.167 / 43.079 \\
5 & -0.40625 & 1.03125 & 1.4375  & 8 & 36.901 / 40.277 & 37.026 / \textbf{40.576} & 35.590 / 39.469 & \textbf{39.274} / 40.354 \\
%\hline 
\specialrule{.15em}{.2em}{.0em}
\end{tabular}
}
\end{center}
%\vspace{-.8em}
\end{table*}

\subsubsection{Evaluation Dataset 1} \label{sec:eva_data_1}
The High Density Camera Array (HDCA) dataset  
\cite{ziegler2017acquisition}
is a real-world 4D light field dataset captured by a DSLR camera mounted on a high-precision gantry. 
This dataset is composed of eight light fields of size $101\times 21 \times 3976 \times 2652\times 3$ and one light field with a size $99 \times 21 \times 3976 \times 2652 \times 3$. 
Note that these raw light field data can hardly be used for evaluation for two reasons: (i) parallax images in these light fields have black borders due to calibration (see \figref{fig:dataset}\,(j)); 
(ii) $d_{range}$ between neighboring views is up to around $5$ pixels, suggesting that these light fields are very SSLFs that can not provide enough ground-truth data for the performance evaluation of the DSLF reconstruction approaches.  
Therefore, a cutting and scaling strategy is proposed to tailor this dataset for the evaluation purpose. 
Specifically, the top 97 horizontal-parallax images of each light field are processed by the cutting operation represented by the 16\,:\,9 red box in \figref{fig:dataset}\,(j) and then downscaled to a new resolution, \ie $1280\times 720$ pixels, using the cubic spline kernel
\cite{gotchev03}. 
Consequently, the evaluation dataset 1 is composed of nine horizontal-parallax ground-truth light fields ${\Psi}_{e}^1$ ($1 \leqslant e \leqslant 9$) with the same angular and spatial resolutions, \ie $\ddot{n}^1=97$, $l^1 = 720$ and $m^1 = 1280$. 
The middle views of these nine ground-truth light fields are shown in \figref{fig:dataset}\,(a)\,-\,(i).
The interpolation rate $\delta^1$ for uniformly sampling an input SSLF $\mathcal{S}_e^1$ from a ground-truth light field ${\Psi}_{e}^1$ is set to 8, 
such that nine input SSLFs $\mathcal{S}_e^1$ ($1 \leqslant e \leqslant 9$) are generated with the same angular resolution $n^1= \big(1+\frac{\ddot{n}^1-1}{\delta^1} \big)= 13$. 
The disparity information of each $\mathcal{S}_e^1$ is first estimated automatically using a state-of-the-art optical flow method, PWC-Net
\cite{sun2018cvpr}, and then refined manually. 
%In particular, for the disparity estimation of $\mathcal{S}_e^1$, $\delta^1$ is first set to $16$ to pick $n^1=7$ parallax images with large disparity ranges from ${\Psi}_{e}^1$. 
%One of the state-of-the-art optical flow methods, \ie PWC-Net
%\cite{sun2018cvpr}, is then applied to these images to approximate the $d_{min}$ and $d_{range}$, which are then double checked manually with a measurement resolution of one pixel because this algorithm may not be accurate enough for estimating these two values. 
%Finally, the estimated $d_{min}$ and $d_{range}$ are divided by 2 for each $\mathcal{S}_e^1$. 
The final approximated $d_{min}^\mathcal{S}$, $d_{max}^\mathcal{S}$ and $d_{range}^\mathcal{S}$ for all the input SSLFs are exhibited in the left part of \tabref{tab:res1}, where $d_{range}^\mathcal{S}$ varies from 8.5 to 15 pixels,
satisfying the aforementioned requirement (b).
%indicating that all the input SSLFs in the evaluation dataset one can be handled by both ST and DRST using the same shearlet system ($\xi = 4$, $\gamma = 127$) because their $d_{range}^\mathcal{S}$ values are less than the sampling interval $\tau=16$.
%For the sampling interval $\tau^e$ at the evaluation stage, $\tau^e \geqslant \lceil d_{range}^e\rceil$ as depicted in  \secref{sec:intro}.
%There is an additional requirement for it, \ie $\tau^e\%\delta^e = 0$, which is because the ground-truth views in ${\Psi}_e$ should have the same spatial positions (at the axis `s' in \figref{fig:intro}\,(a)) as the corresponding reconstructed views in the target DSLF $\mathcal{D}_e$.
%As explained in \secref{sec:dcst}, the shearlet system of DRST is constructed for DSLF reconstruction on SSLFs with varying moderate disparity ranges, \ie $\tau^e \leqslant 16$ pixels. 
%Therefore, $\tau^e$ is set to a fixed value of $16$ pixels for all evaluation SSLFs $\mathcal{S}_e$ ($1 \leqslant e \leqslant 9$). 
The target DSLF $\mathcal{D}_e^1$ to be reconstructed from an input SSLF $\mathcal{S}_e^1$ consists of $\dot{n}^1=\big((n^1-1)\tau+1\big)=193$ horizontal-parallax images.

\begin{figure*}[t]

\begin{minipage}[t]{0.195\linewidth}
  \centering
  \centerline{\includegraphics[width=1.\textwidth]{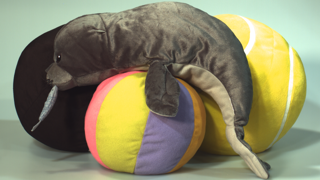}}
  \vspace{-.4em}
  \centerline{\scriptsize(a) $\Psi_1^3$: Seal and balls}\medskip
\end{minipage}
\hfill
\begin{minipage}[t]{0.195\linewidth}
  \centering
  \centerline{\includegraphics[width=1.\textwidth]{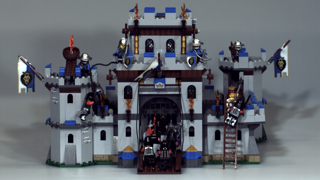}}
  \vspace{-.4em}
  \centerline{\scriptsize(b) $\Psi_2^3$: Castle}\medskip
\end{minipage}
\hfill
\begin{minipage}[t]{0.195\linewidth}
  \centering
  \centerline{\includegraphics[width=1.\textwidth]{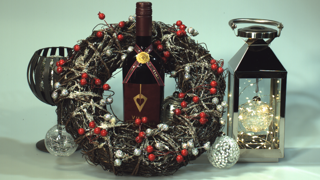}}
  \vspace{-.4em}
  \centerline{\scriptsize(c) $\Psi_3^3$: Holiday}\medskip
\end{minipage}
\hfill
\begin{minipage}[t]{0.195\linewidth}
  \centering
  \centerline{\includegraphics[width=1.\textwidth]{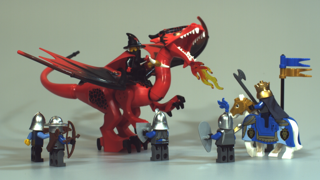}}
  \vspace{-.4em}
  \centerline{\scriptsize(d) $\Psi_4^3$: Dragon}\medskip
\end{minipage}
\hfill
\begin{minipage}[t]{0.195\linewidth}
  \centering
  \centerline{\includegraphics[width=1.\textwidth]{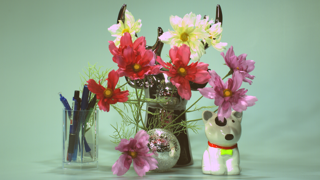}}
  \vspace{-.4em}
  \centerline{\scriptsize(f) $\Psi_5^3$: Flowers}\medskip
\end{minipage}
 
\vspace{-.8em}
\caption{Middle views of ground-truth light fields ${\Psi}_e^3$ ($1 \leqslant e \leqslant 5$) in the evaluation dataset 3.}
%\vspace{.4em}
\label{fig:dataset3}
\end{figure*}

\begin{table*}[!t]
\begin{center}
\caption{Disparity estimation, minimum and average per-view PSNR results (in dB, explained in \secref{sec:qua}) for the performance evaluation of different light field reconstruction methods on the evaluation dataset 3. %evaluation datasets $\mathcal{S}_e$ ($1 \leqslant e \leqslant 9$). 
} 
\label{tab:res3}
\vspace{-.8em}
\scalebox{1.}{
\begin{tabular}{c||c|c|c|c|c|c}
\specialrule{.15em}{.0em}{.2em}
\multirow{2}{*}{$e$ of $\mathcal{S}_e^3$} & \multicolumn{3}{c|}{\textbf{Disparity Estimation (pixel)} } & \multicolumn{3}{c}
{\textbf{Minimum PSNR / Average PSNR (dB)} } \\ [.5ex] 
\hhline{|~||------} 
  & $d_{min}^\mathcal{S}$ & $d_{max}^\mathcal{S}$ & $d_{range}^\mathcal{S}$  & SepConv ($\mathcal{L}_1$) \cite{niklaus2017iccv} & ST \cite{vagharshakyan2017accelerated} & DRST 
  \\ [.1ex] 
  \hline
1 & -10.5 & 3.5 & 14  & 41.619 / 43.194 & 37.473 / 42.883 & \textbf{43.051} / \textbf{44.080} \\
2 & -2    & 12  & 14  & \textbf{35.256} / \textbf{37.167} & 34.719 / 36.969 & 35.253 / 37.141 \\
3 & -8    & 6   & 14  & \textbf{30.884} / \textbf{34.709} & 30.428 / 34.017 & 30.631 / 34.363 \\
4 & -9    & 7   & 16  & 41.061 / 41.808 & 38.087 / 41.876 & \textbf{41.389} / \textbf{42.429} \\
5 & -6.5  & 7.5 & 14  & 36.018 / \textbf{37.994} & 34.851 / 37.787 & \textbf{36.186} / 37.857 \\
%\hline 
\specialrule{.15em}{.2em}{.0em}
\end{tabular}
}
\end{center}
%\vspace{-.8em}
\end{table*}

%& PIASC ($\mathcal{L}_1$) \cite{gao2018icmew}
%& \textbf{43.091} / \textbf{44.215}
%& \textbf{35.491} / \textbf{37.419}
%& \textbf{30.974} / \textbf{34.842}
%& \textbf{41.561} / \textbf{42.559}
%& 36.104 / \textbf{38.062}         

\subsubsection{Evaluation Dataset 2}	% MPI
The MPI light field archive contains five real-world horizontal-parallax light fields captured by  one-meter long motorized linear stage
\cite{kiran2017towards}.
Each source 3D light field is composed of 101 horizontal-parallax images of the same resolution, \ie $960\times720$ pixels. 
Following the same dataset preparation process as above, the top 97 images are chosen to form a ground-truth light field from each source 3D light field.
Therefore, the evaluation dataset 2 has five horizontal-parallax ground-truth light fields ${\Psi}_{e}^2$ ($1 \leqslant e \leqslant 5$) with the same angular and spatial resolutions, \ie $\ddot{n}^2=97$, $l^2=720$ and $m^2=960$.
The middle views of these five horizontal-parallax ground-truth light fields are exhibited in \figref{fig:dataset2}.
Regarding the disparity estimation of these five ground-truth 3D light fields, the interpolation rate $\delta$ is first set to 32 to generate five SSLFs with $n=7$ parallax images having large disparities. 
The $d_{min}$ and $d_{max}$ of these five generated SSLFs are then estimated by hands with one-pixel measurement resolution. 
Finally, these estimated $d_{min}$ and $d_{max}$ are divided by $\delta=32$ to produce the final disparity estimations of the ground-truth light fields ${\Psi}_{e}^2$ ($1 \leqslant e \leqslant 5$), 
which are shown in the left part of \tabref{tab:res2}.
It can be seen that the $d_{range}^\Psi$ value of $\Psi_1^2$ is above $2$ pixels, the $d_{range}^\Psi$ values of $\Psi_2^2$ and $\Psi_5^2$ are between $1$\,-\,$2$ pixels, and the $d_{range}^\Psi$ values of $\Psi_3^2$ and $\Psi_4^2$ are less than $1$ pixel.
Since the baseline approaches (SepConv and PIASC) require the interpolation rate $\delta^2$ to be a power of two and DRST requires that $ d_{range}^\mathcal{S} = (d_{range}^\Psi\cdot\delta^2) \leqslant \tau$ ($=16$) for any input SSLF,
the interpolation rate $\delta^2$ is set to 4 for $\Psi_1^2$, 
8 for $\Psi_2^2$ and $\Psi_5^2$, 
and 16 for $\Psi_3^2$ and $\Psi_4^2$, respectively.  
Therefore, five input SSLFs $\mathcal{S}_e^2$ ($1 \leqslant e \leqslant 5$) are generated. 
Specifically, $\mathcal{S}_1^2$ has $n^2=25$ parallax views,  
$\mathcal{S}_2^2$ and $\mathcal{S}_5^2$ have $n^2=13$ parallax views, and $\mathcal{S}_3^2$ and $\mathcal{S}_4^2$ have $n^2=7$ parallax views.
The target DSLFs to be reconstructed from these five input SSLFs have varying angular resolutions. 
To be precise, $\mathcal{D}_1^2$ has $\dot{n}^2=385$ parallax views, $\mathcal{D}_2^2$ and $\mathcal{D}_5^2$ have $\dot{n}^2=193$ parallax views, and $\mathcal{D}_3^2$ and $\mathcal{D}_4^2$ have $\dot{n}^2=97$ parallax views.

\subsubsection{Evaluation Dataset 3} 	% TUT
The evaluation dataset 3 is Centre for Immersive Visual Technologies (CIVIT) DSLF dataset, which was prepared for IEEE International Conference on Multimedia and Expo (ICME) 2018 grand challenge on DSLF reconstruction
\cite{civit, gao2018icmew}.
The dataset has five real-world horizontal-parallax light fields.
In particular, the evaluation dataset 3 contains five ground-truth horizontal-parallax light fields $\Psi_e^3$ ($1 \leqslant e \leqslant 5$) with the same angular and spatial resolutions, \ie $\ddot{n}^3=193$, $l^3=720$ and $m^3=1280$.
The middle views of these ground-truth light fields are illustrated in \figref{fig:dataset3}.
The input SSLFs $\mathcal{S}_e^3$ ($1 \leqslant e \leqslant 5$) of the evaluation dataset 3 are generated from $\Psi_e^3$ ($1 \leqslant e \leqslant 5$) using the interpolation rate $\delta^3=16$. 
Each generated SSLF has $n^3=13$ parallax images accordingly.
%For the disparity estimation of the input SSLFs, the same disparity estimation strategy as above is performed here. 
%Specifically, the interpolation rate $\delta^3$ is first set to $32$ to produce five SSLFs with $n^3=7$ large-disparity-range parallax images. 
%The $d_{min}$ and $d_{max}$ of them are estimated manually with the measurement resolution of one pixel.
%After dividing the estimated $d_{min}$ and $d_{max}$ values by 2, 
The disparity data of all the input SSLFs are estimated manually and exhibited in the left part of \tabref{tab:res3}.
It can be seen that the estimated $d_{range}^\mathcal{S}$ values of the input SSLFs %are less than or equal to the sampling interval $\tau$ ($=16$). %except for the $d_{range}^\mathcal{S}$ of $\mathcal{S}_4^3$, 
meet the aforementioned requirement (b).
%which is mainly caused by the measurement resolution of the manual way. 
The target DSLFs $\mathcal{D}_e^3$ ($1 \leqslant e \leqslant 5$) to be reconstructed from the input SSLFs have the same angular resolution as the ground-truth light fields $\Psi_e^3$ ($1 \leqslant e \leqslant 5$), \ie $\dot{n}^3=\ddot{n}^3=193$.

%\noindent\textbf{Evaluation criteria.}

%It is worth to be mentioned that all the ground-truth light fields $\Psi_e$ ($1 \leqslant e \leqslant 9$) are not densely-sampled because $\frac{d_{range}^e}{\delta} \geqslant \frac{d_{range}^7}{\delta} = 1.125$ pixels. 
%Therefore, the per-view PSNR is calculated \textit{w.r.t.} recovering each ``unknown'' view in $\Psi_e \setminus \mathcal{S}_e$, of which the view number is  $(\ddot{n}^e - n^e) = 84$ in total. 
%However, each reconstructed $\mathcal{D}_e$ contains $(\dot{n}^e - \ddot{n}^e) = 96$ redundant views that are not used for per-view PSNR calculations, which is because the interpolation rate $\delta^e$ ($=8$) for $\mathcal{S}_e$  is only half of the sampling interval $\tau^e$ ($=16$) of $\mathcal{D}_e$.
%\cite{gao2019mast}. 

\subsubsection{Implementation details} 
%At the training stage, each pre-sheared and padded sparsely-sampled EPI $\varepsilon'_i$ has the same resolution of $672\times 172$ ($m' \times n'$) pixels and each mini-batch has four $(\varepsilon'_i,\, \theta_i,\, \theta'_i)$. 
The proposed DRST approach is implemented using TensorFlow\,2\footnote{\url{https://github.com/ygaostu/DRST} (to appear)} and trained on a server with an Nvidia GeForce RTX 2080\,Ti GPU for ten epochs.
The optimization tool for minimizing the loss function \eqref{eq:loss} is AdaMax 
\cite{kingma2014adam} with parameters $\beta_1=0.9$ and $\beta_2=0.999$.
In addition, the learning rate is set to $10^{-3}$ for the first two epochs and then adjusted to $10^{-4}$ for the rest eight epochs.  
A mini-batch is composed of four sparsely-sampled EPIs $\hat{\varepsilon}$ that are extracted from $\ddot{\varepsilon}$ as described in \secref{sec:pre-sh-pad} (see \figref{fig:train}\,(d) and (c)).
Considering that each $\hat{\varepsilon}$ has three color channels, a mini-batch actually comprises $12$ one-channel EPIs.
Since the number of training samples is given above in \secref{sec:train_data}, each epoch has 124,416 training iterations.
The encoder-decoder network $\mathcal{R}$ in \secref{sec:sparsity_regu} has $3,618,959$ trainable parameters. 
It takes around 32 hours to finish the whole training process. 
Regarding the evaluation on the above three evaluation datasets, all the methods are conducted on a local machine with an Nvidia GeForce RTX 2070 GPU. 
It should be noted that when evaluating DRST on an input evaluation SSLF with an angular resolution  $n$ ($>3$),
this SSLF requires to be converted into $\lfloor \frac{n}{2} \rfloor$ sub-SSLFs, of which each has the same angular resolution, \ie 3 pixels.  
%The sparsely-sampled EPIs in each sub-SSLF can be transformed into the sparsely-sampled EPI $\tilde{\varepsilon}$ via pre-shearing and zero-padding.
The parameters of ST using the DORE algorithm are set in accordance with
\cite{vagharshakyan2017accelerated}, 
\ie $\alpha = 20$ with 100 iterations and a low-pass initial estimation for each input sparsely-sampled EPI.

%At the evaluation stage, each processed sparsely-sampled EPI $\varepsilon'_i$ is in the size of  $1312 \times 128$ ($m'\times n'$) pixels. 
%Note that the evaluation EPI $\varepsilon'_i$ is different from the $\varepsilon''_i$ introduced in \secref{sec:dcst} for DSLF training. 
%Specifically, each $\varepsilon'_i$ here has only three color lines arranged in the same way as \figref{fig:pipeline}\,(a). 
%This means that the prediction function of DRST at the evaluation stage is equal to synthesizing $2(\tau^e-1)=30$ views on three neighboring views in $\mathcal{S}_e$ and this procedure is repeated $\lfloor \frac{n^e}{2} \rfloor = 6$ times until the target $\mathcal{D}_e$ is completely reconstructed. 

\begin{figure*}[t]
\begin{minipage}[t]{0.23\linewidth}
  \centering
  \centerline{\includegraphics[width=1.\textwidth]{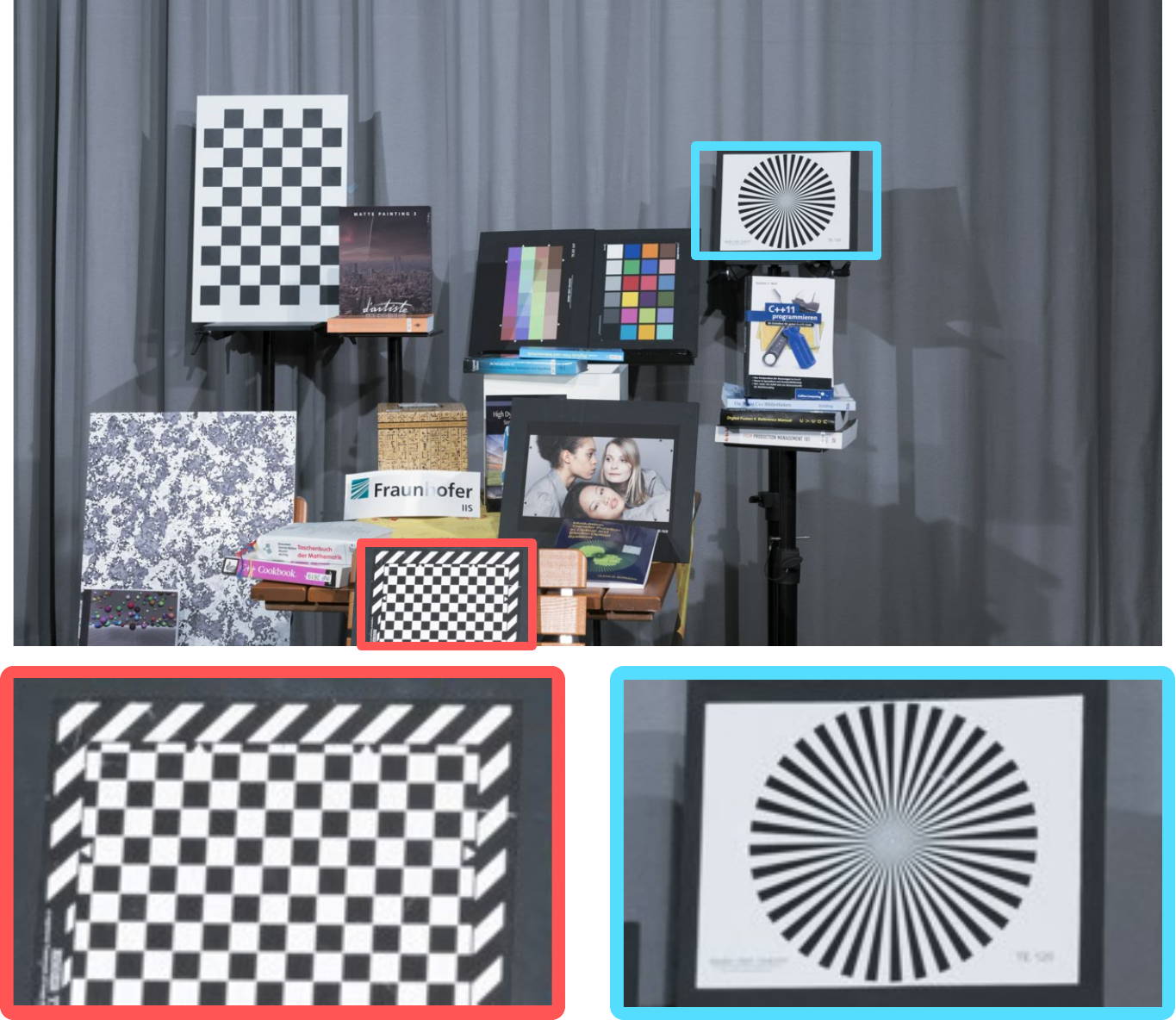}}
  \vspace{-.4em}
  \centerline{\scriptsize(a) $\mathcal{I}_{93}$ of $\Psi_1^1$ (Ground-truth)}\medskip
\end{minipage}
\hfill
\begin{minipage}[t]{0.23\linewidth}
  \centering
  \centerline{\includegraphics[width=1.\textwidth]{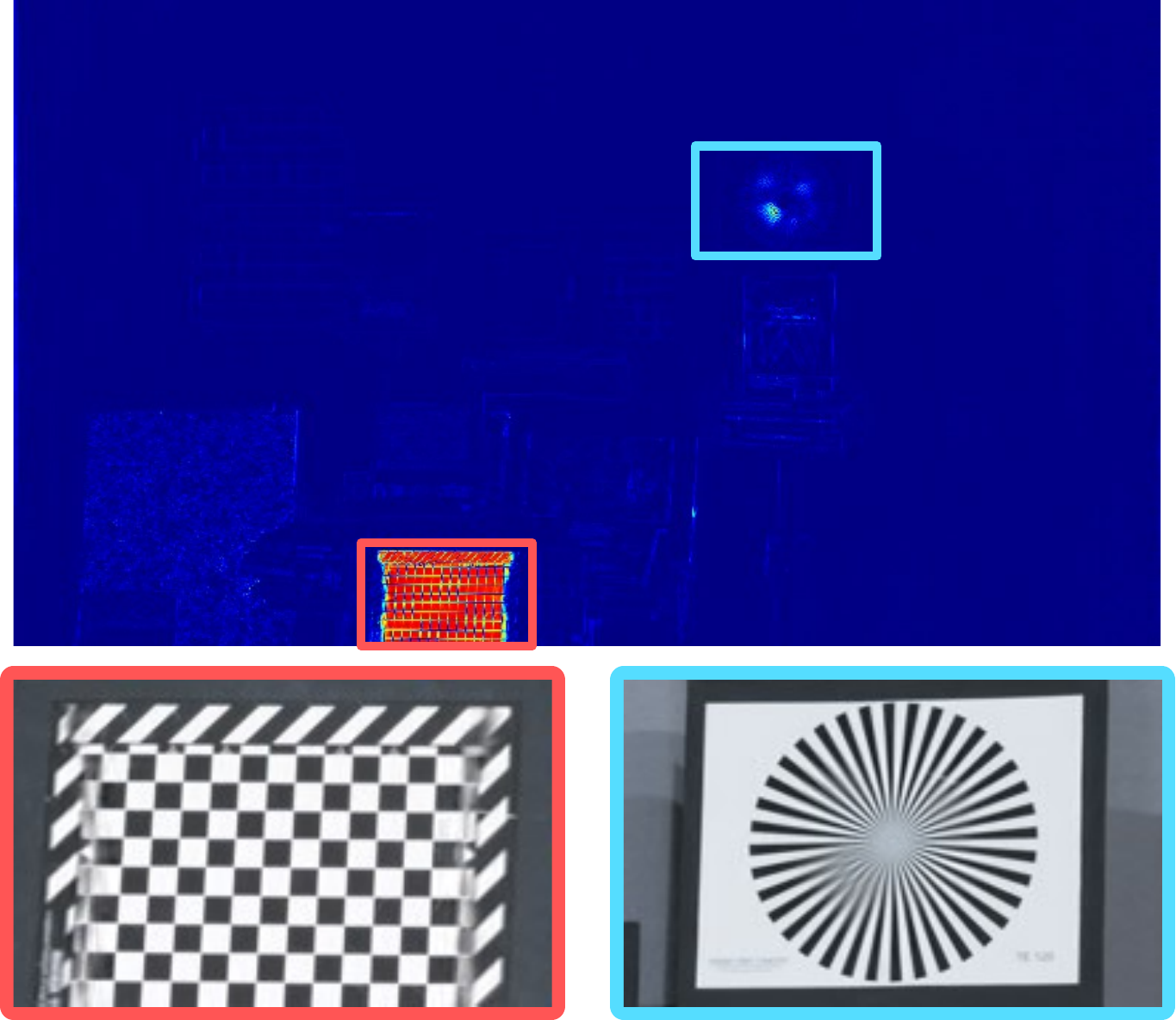}}
  \vspace{-.4em}
  \centerline{\scriptsize(b) PIASC ($\mathcal{L}_1$) \cite{gao2018icmew} (22.020 dB)}\medskip
\end{minipage}
\hfill
\begin{minipage}[t]{0.23\linewidth}
  \centering
  \centerline{\includegraphics[width=1.\textwidth]{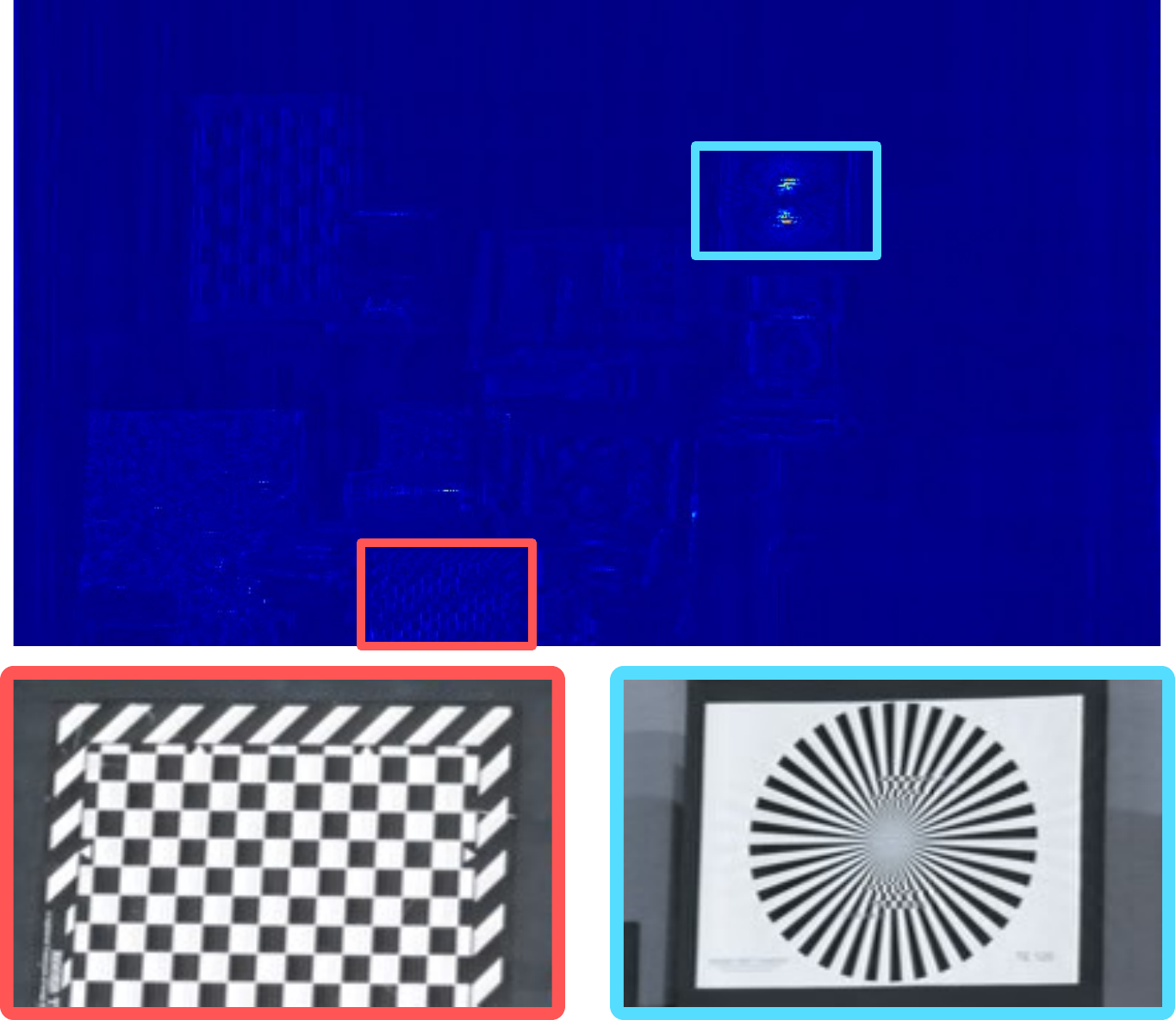}}
  \vspace{-.4em}
  \centerline{\scriptsize(c) ST \cite{vagharshakyan2017accelerated} (35.660 dB)}\medskip
\end{minipage}
\hfill
\begin{minipage}[t]{0.251904762\linewidth}%0.2402
  \centering
  \centerline{\includegraphics[width=1.\textwidth]{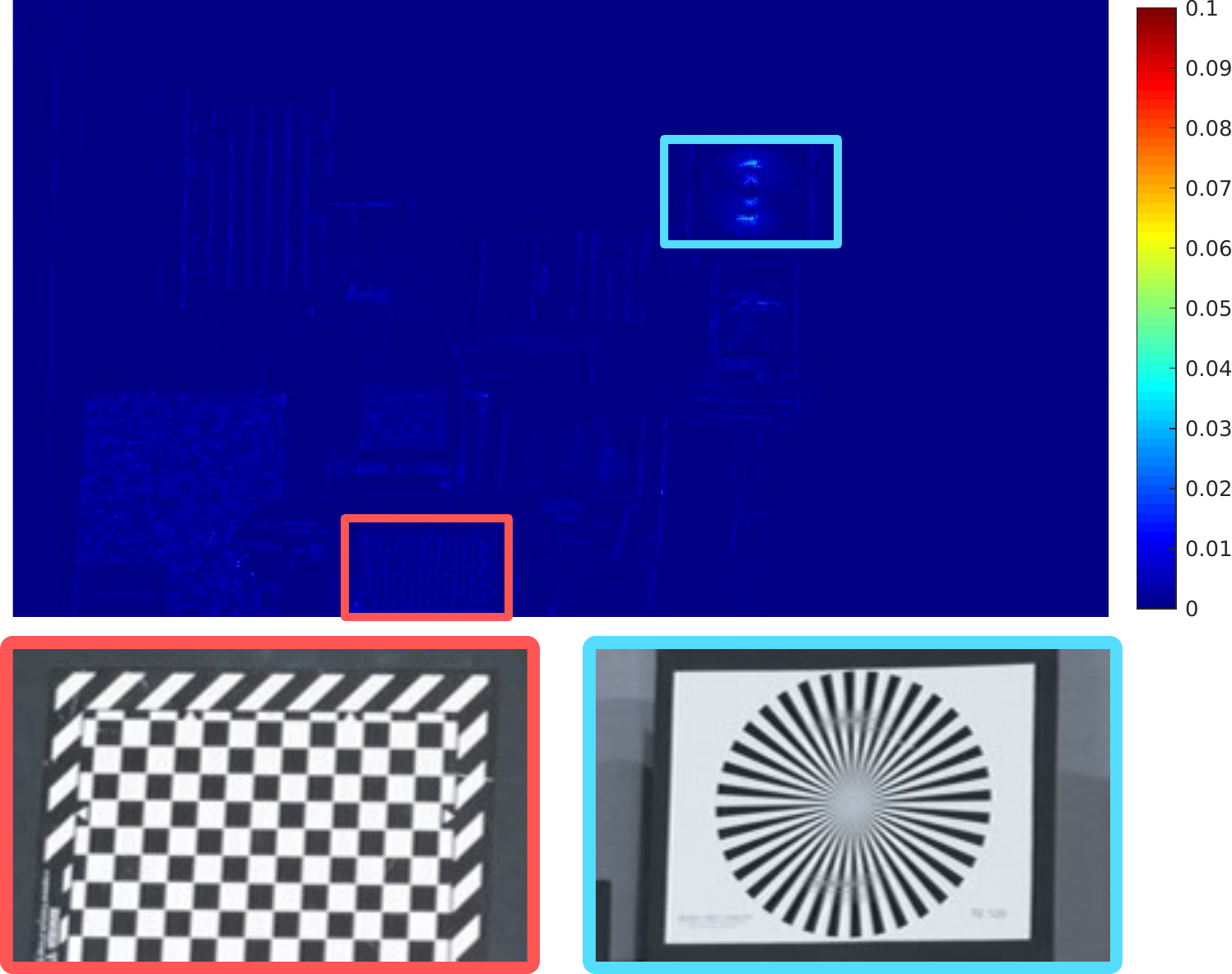}}
  \vspace{-.4em}
  \centerline{\scriptsize(d) DRST (39.008 dB)}\medskip
\end{minipage}
%\hfill

\vspace{-.3em}

%\hfill
\begin{minipage}[t]{0.23\linewidth}
  \centering
  \centerline{\includegraphics[width=1.\textwidth]{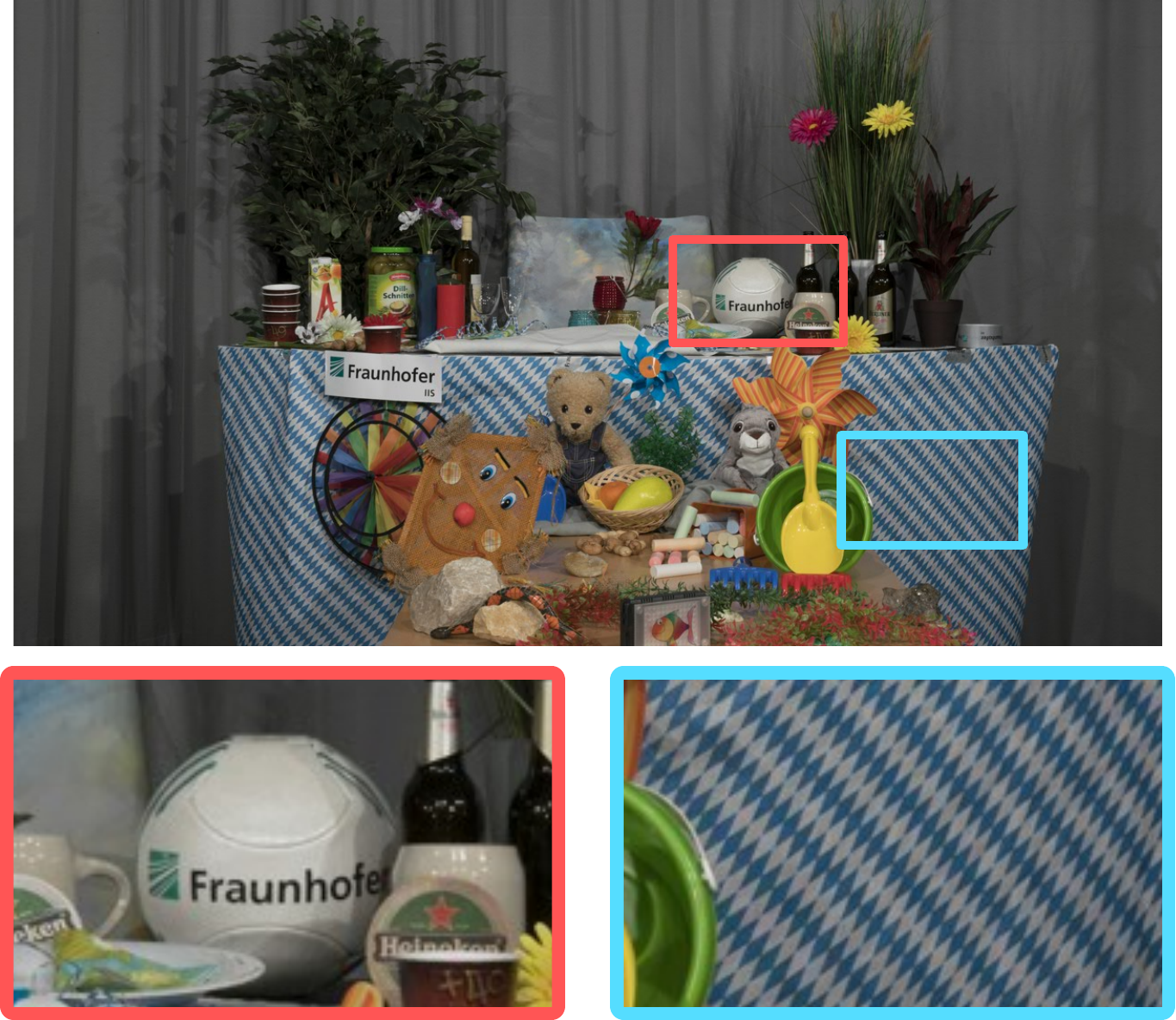}}
  \vspace{-.4em}
  \centerline{\scriptsize(e) $\mathcal{I}_{5}$ of $\Psi_6^1$ (Ground-truth)}\medskip
\end{minipage}
\hfill
\begin{minipage}[t]{0.23\linewidth}
  \centering
  \centerline{\includegraphics[width=1.\textwidth]{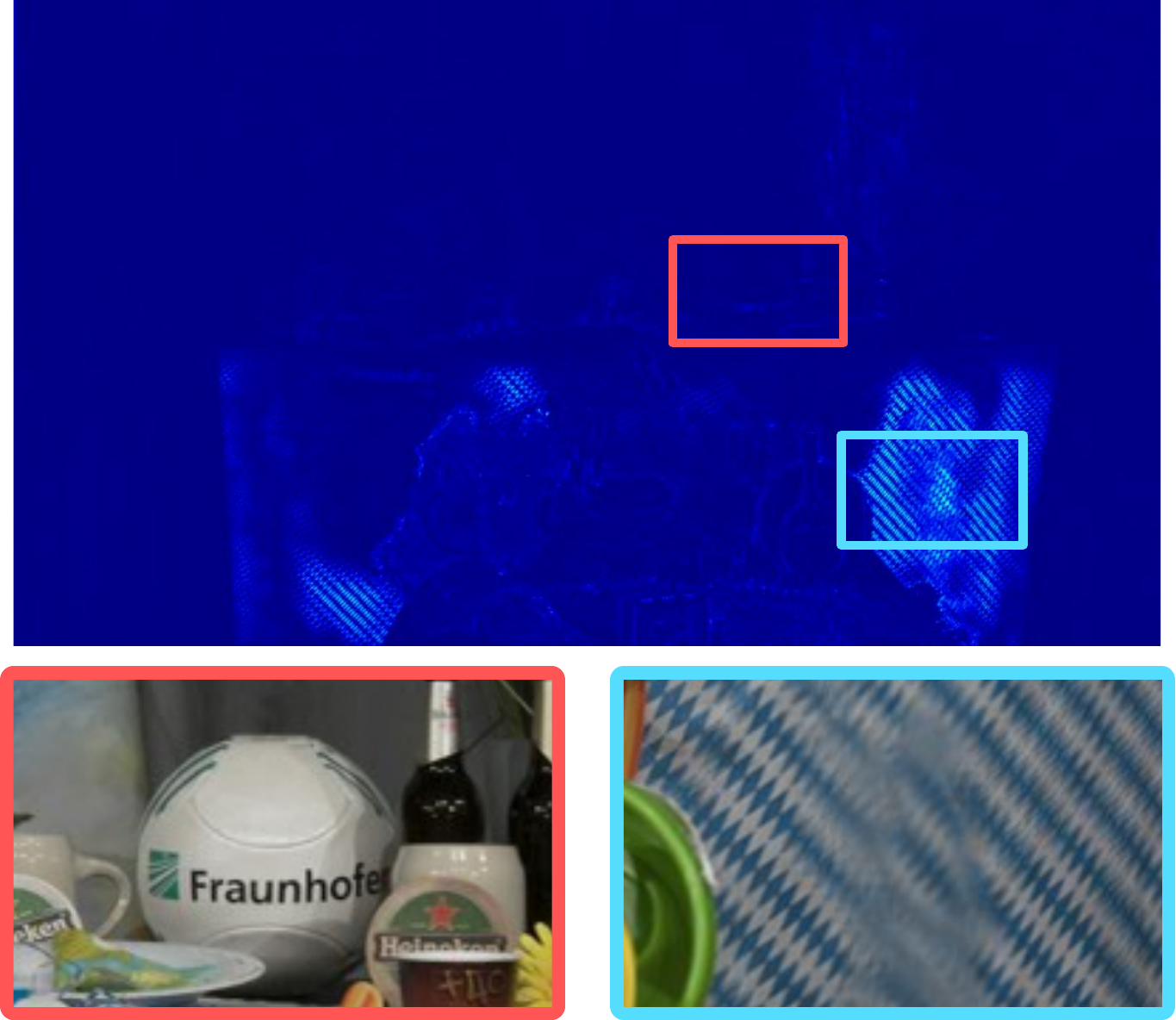}}
  \vspace{-.4em}
  \centerline{\scriptsize(f) PIASC ($\mathcal{L}_1$) \cite{gao2018icmew} (30.914 dB)}\medskip
\end{minipage}
\hfill
\begin{minipage}[t]{0.23\linewidth}
  \centering
  \centerline{\includegraphics[width=1.\textwidth]{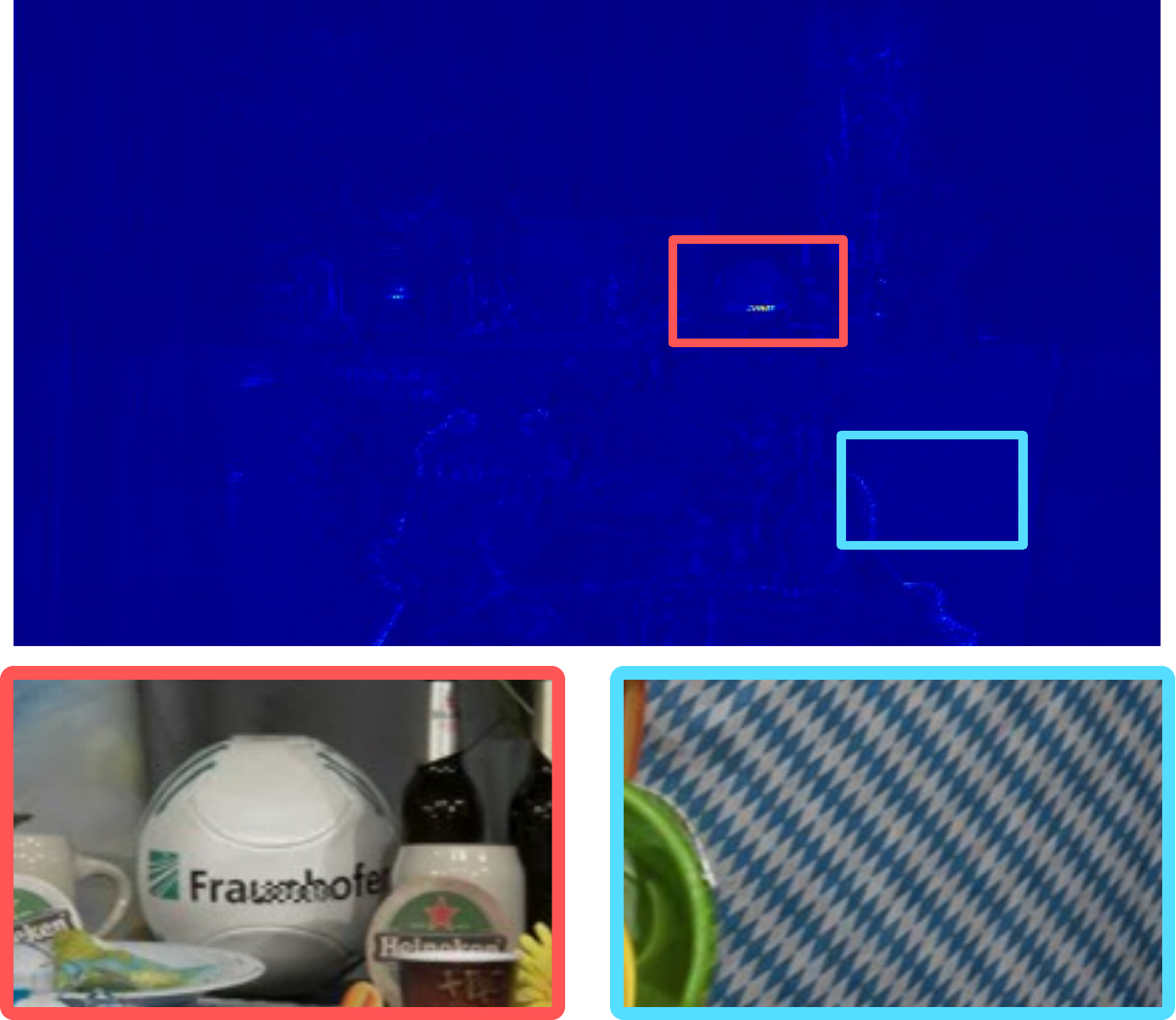}}
  \vspace{-.4em}
  \centerline{\scriptsize(g) ST \cite{vagharshakyan2017accelerated} (37.387 dB)}\medskip
\end{minipage}
\hfill
\begin{minipage}[t]{0.251904762\linewidth}
  \centering
  \centerline{\includegraphics[width=1.\textwidth]{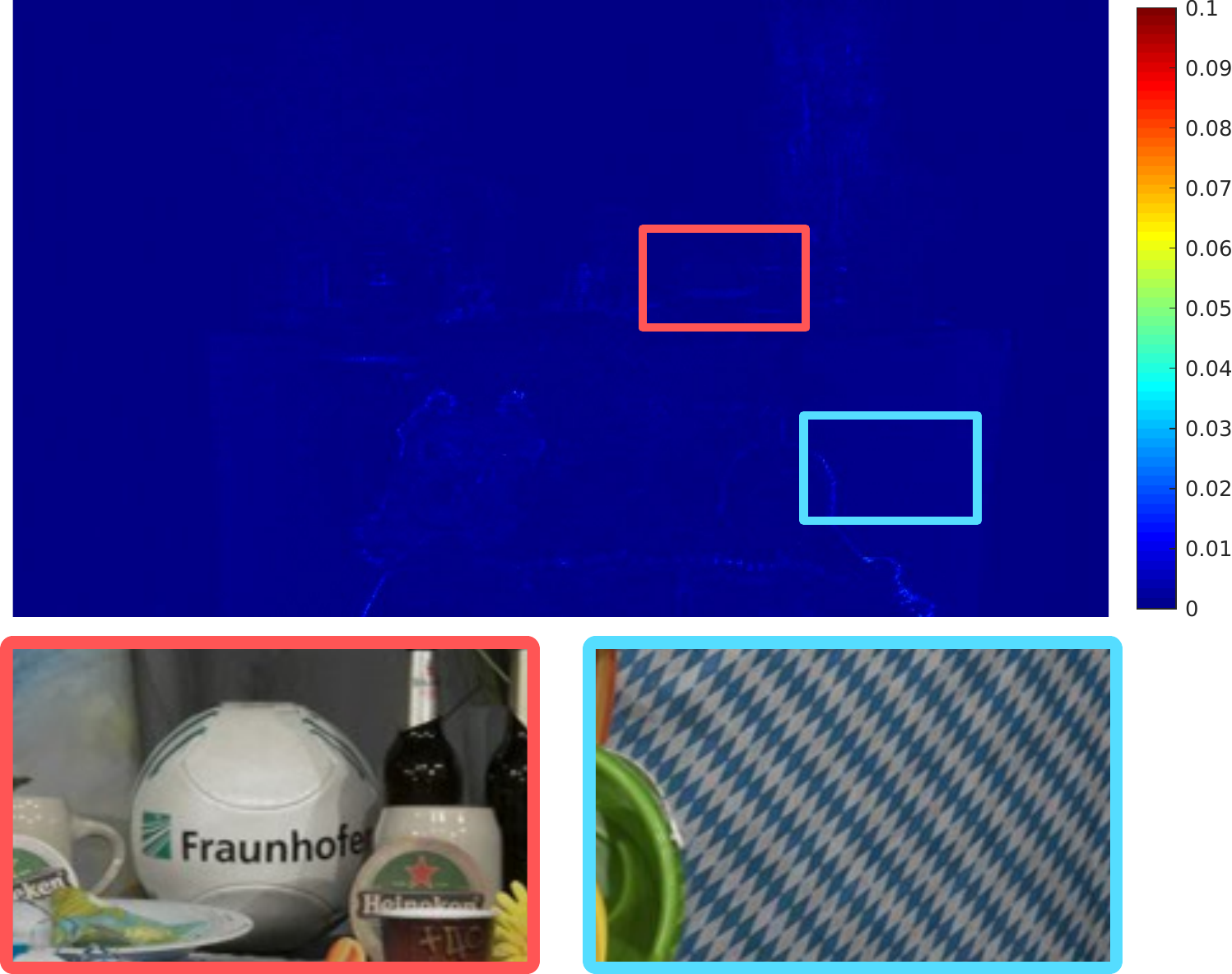}}
  \vspace{-.4em}
  \centerline{\scriptsize(h) DRST (42.654 dB)}\medskip
\end{minipage}
%\hfill

\vspace{-.6em}
\caption{Light field reconstruction results on the evaluation dataset 1.}
\vspace{.2em}
\label{fig:vis1}
\end{figure*}

\begin{figure*}[t]
\begin{minipage}[t]{0.23\linewidth}
  \centering
  \centerline{\includegraphics[width=1.\textwidth]{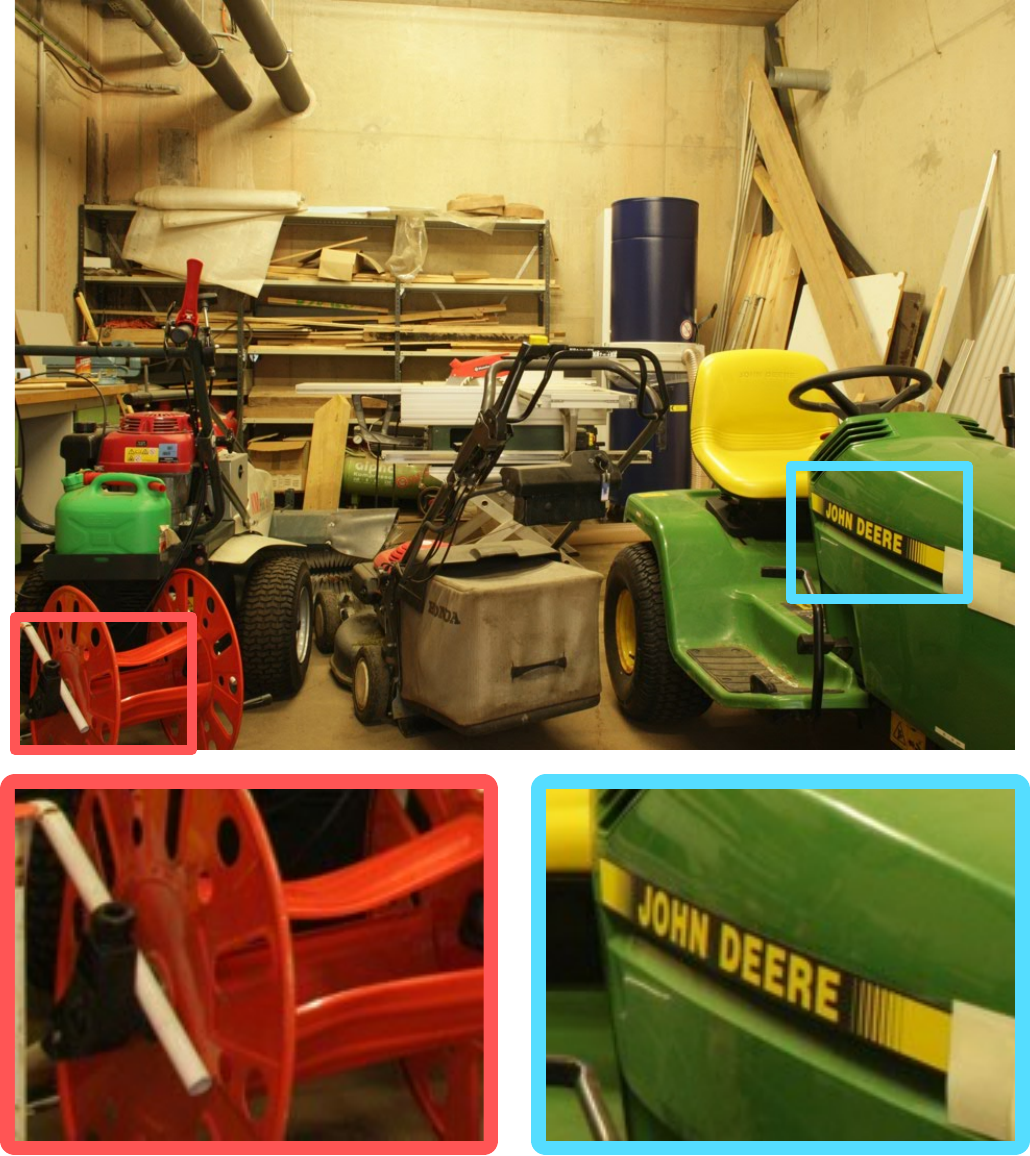}}
  \vspace{-.4em}
  \centerline{\scriptsize(a) $\mathcal{I}_{58}$ of $\Psi_5^2$ (Ground-truth)}\medskip
\end{minipage}
\hfill
\begin{minipage}[t]{0.23\linewidth}
  \centering
  \centerline{\includegraphics[width=1.\textwidth]{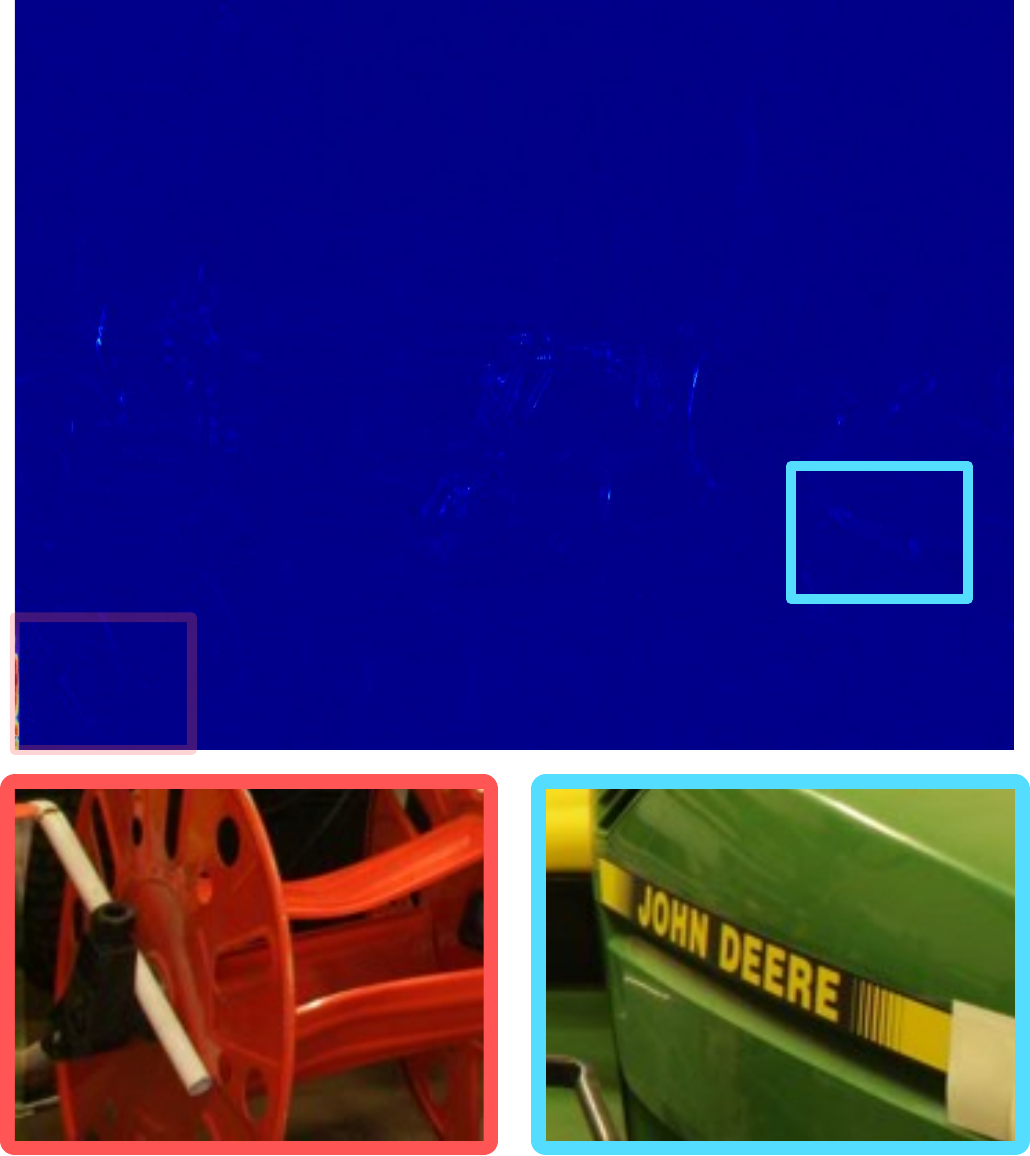}}
  \vspace{-.4em}
  \centerline{\scriptsize(b) PIASC ($\mathcal{L}_1$) \cite{gao2018icmew} (37.026 dB)}\medskip
\end{minipage}
\hfill
\begin{minipage}[t]{0.23\linewidth}
  \centering
  \centerline{\includegraphics[width=1.\textwidth]{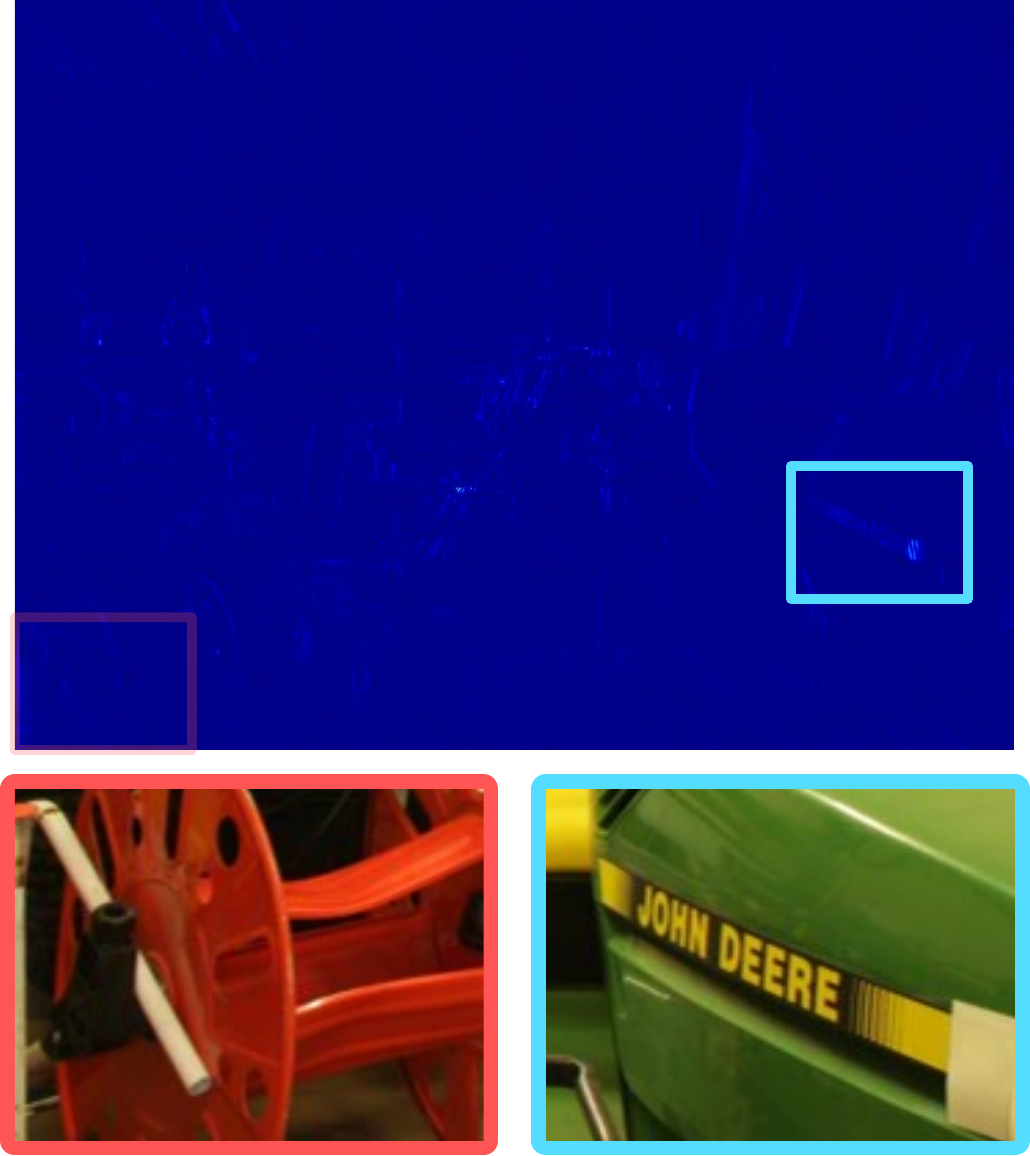}}
  \vspace{-.4em}
  \centerline{\scriptsize(c) ST \cite{vagharshakyan2017accelerated} (40.176 dB)}\medskip
\end{minipage}
\hfill
\begin{minipage}[t]{0.259029126\linewidth}%0.2402
  \centering
  \centerline{\includegraphics[width=1.\textwidth]{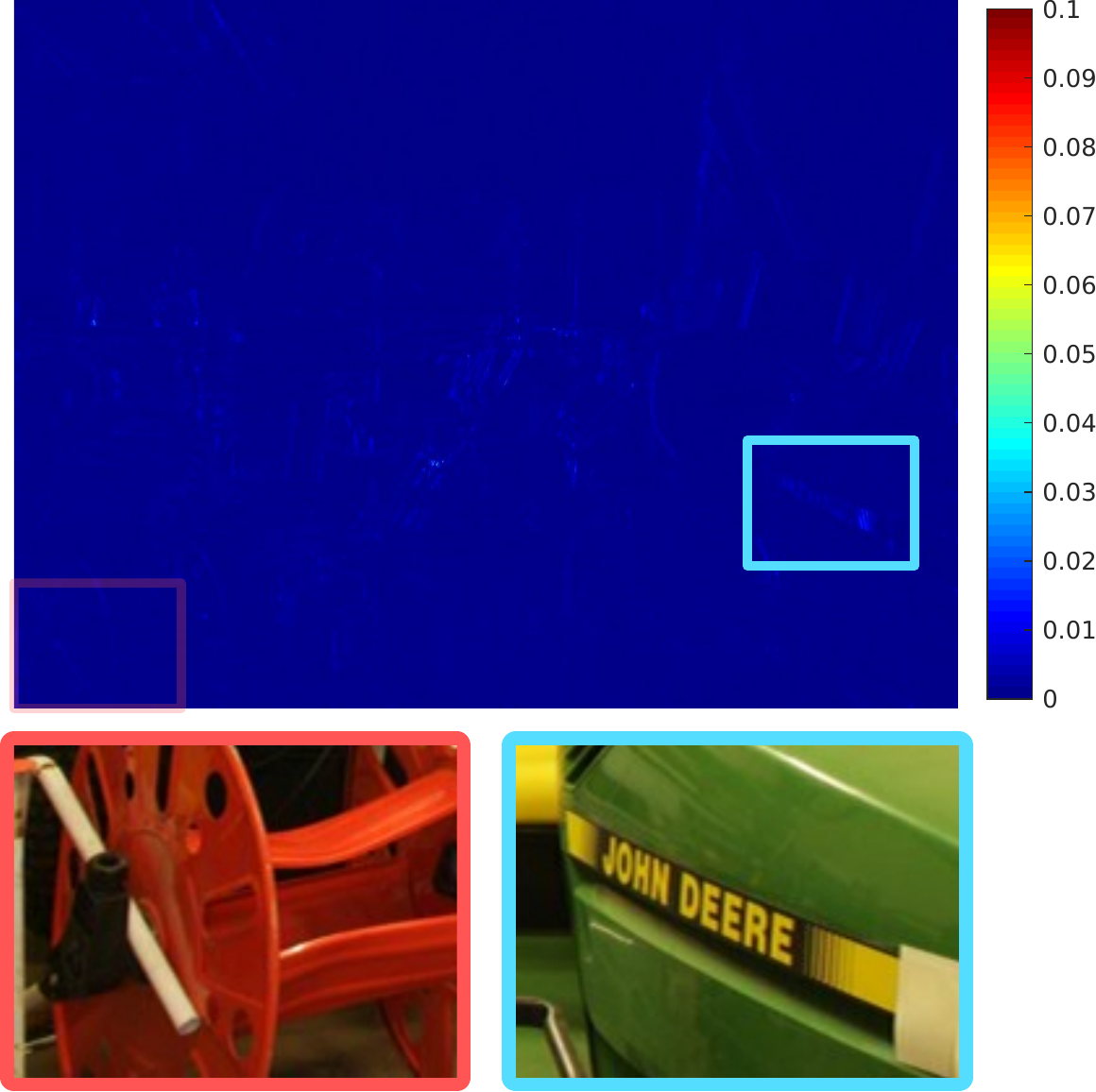}}
  \vspace{-.4em}
  \centerline{\scriptsize(d) DRST (40.471 dB)}\medskip
\end{minipage}
%\hfill

\vspace{-.8em}
\caption{Light field reconstruction results on the evaluation dataset 2.}
\vspace{.2em}
\label{fig:vis2}
\end{figure*}

\begin{figure*}[t]
\begin{minipage}[t]{0.23\linewidth}
  \centering
  \centerline{\includegraphics[width=1.\textwidth]{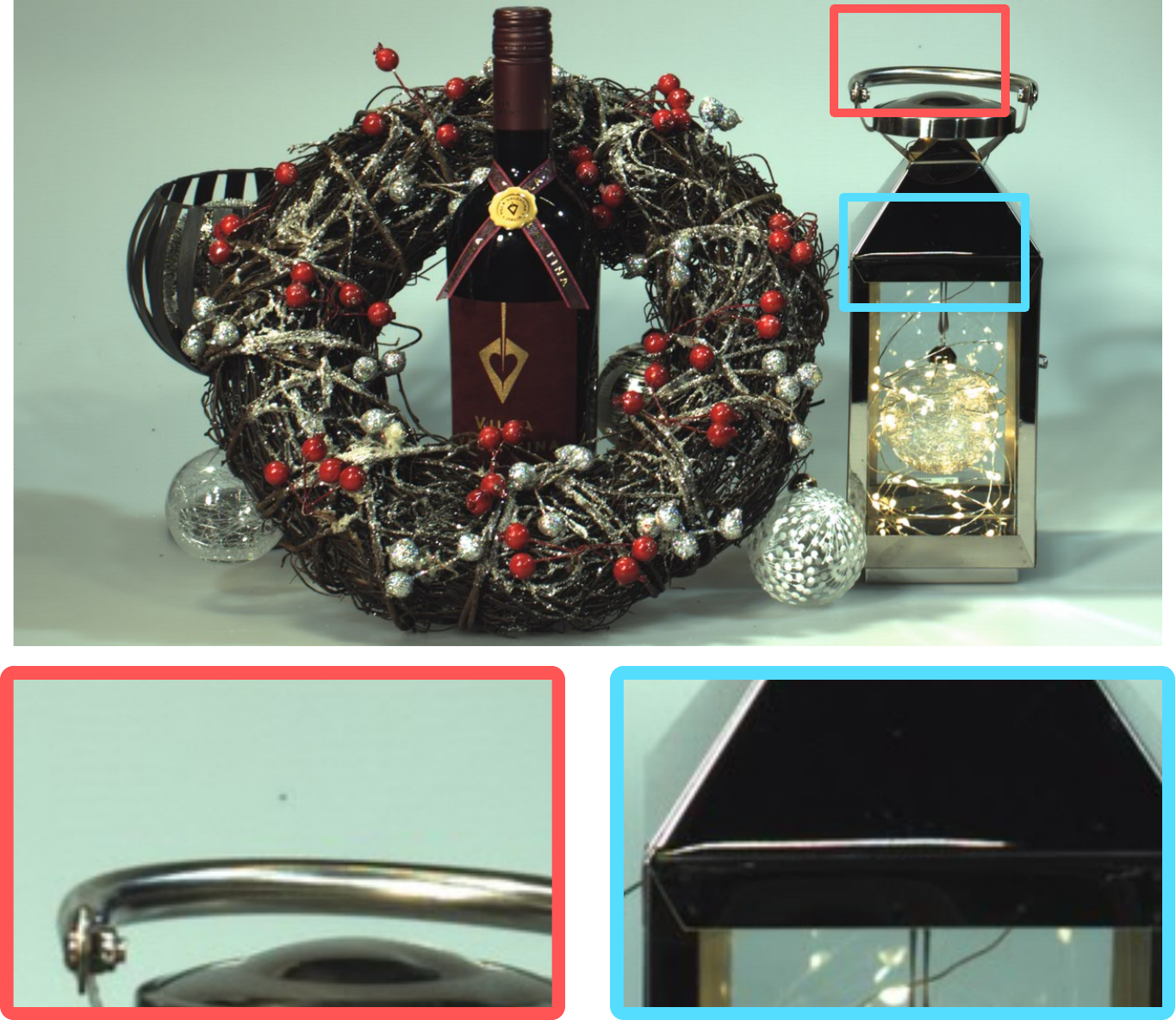}}
  \vspace{-.4em}
  \centerline{\scriptsize(a) $\mathcal{I}_{6}$ of $\Psi_3^3$ (Ground-truth)}\medskip
\end{minipage}
\hfill
\begin{minipage}[t]{0.23\linewidth}
  \centering
  \centerline{\includegraphics[width=1.\textwidth]{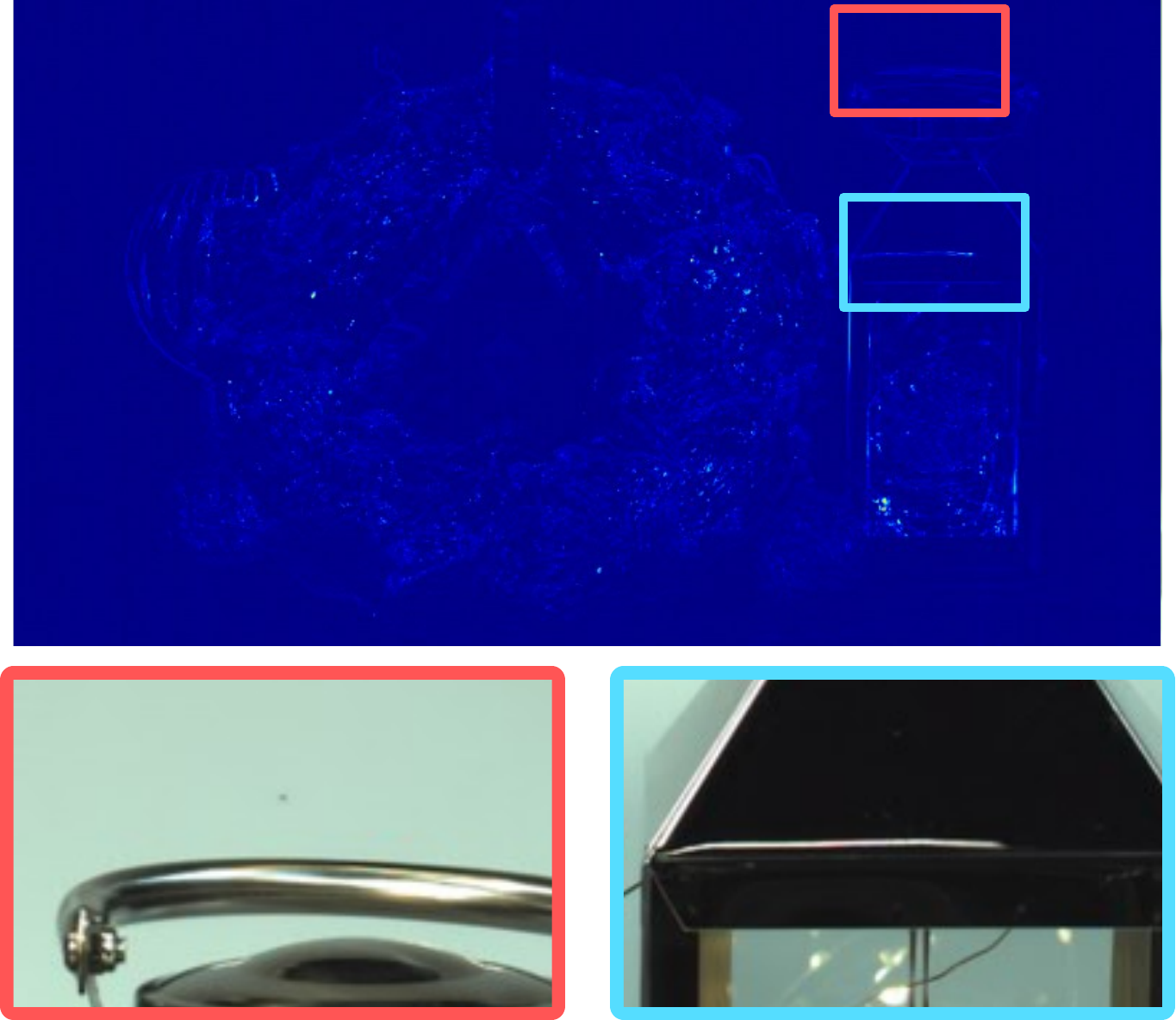}}
  \vspace{-.4em}
  \centerline{\scriptsize(b) SepConv ($\mathcal{L}_1$) \cite{niklaus2017iccv} (33.479 dB)}\medskip %33.672
\end{minipage}
\hfill
\begin{minipage}[t]{0.23\linewidth}
  \centering
  \centerline{\includegraphics[width=1.\textwidth]{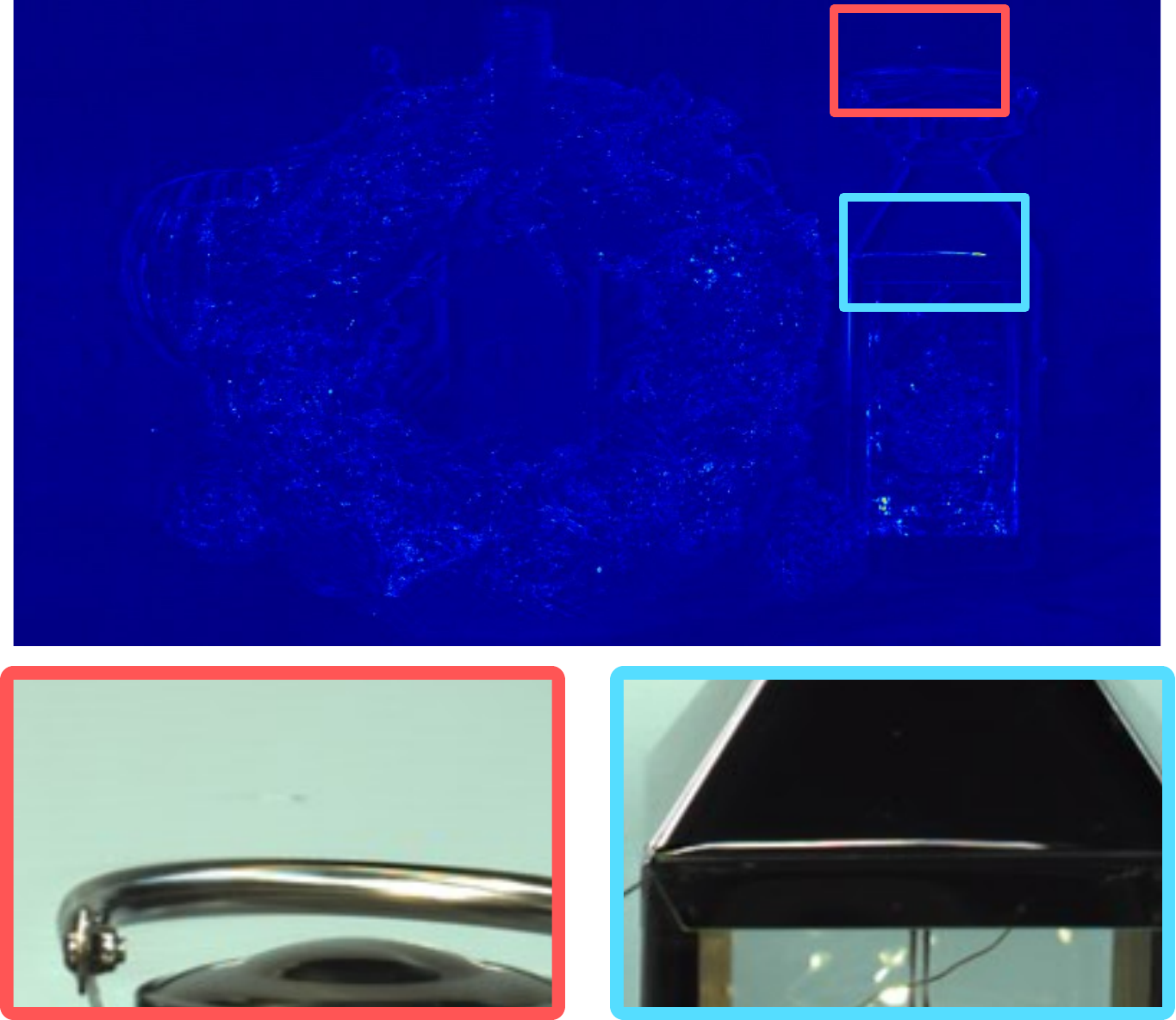}}
  \vspace{-.4em}
  \centerline{\scriptsize(c) ST \cite{vagharshakyan2017accelerated} (31.315 dB)}\medskip
\end{minipage}
\hfill
\begin{minipage}[t]{0.251904762\linewidth}%0.2402
  \centering
  \centerline{\includegraphics[width=1.\textwidth]{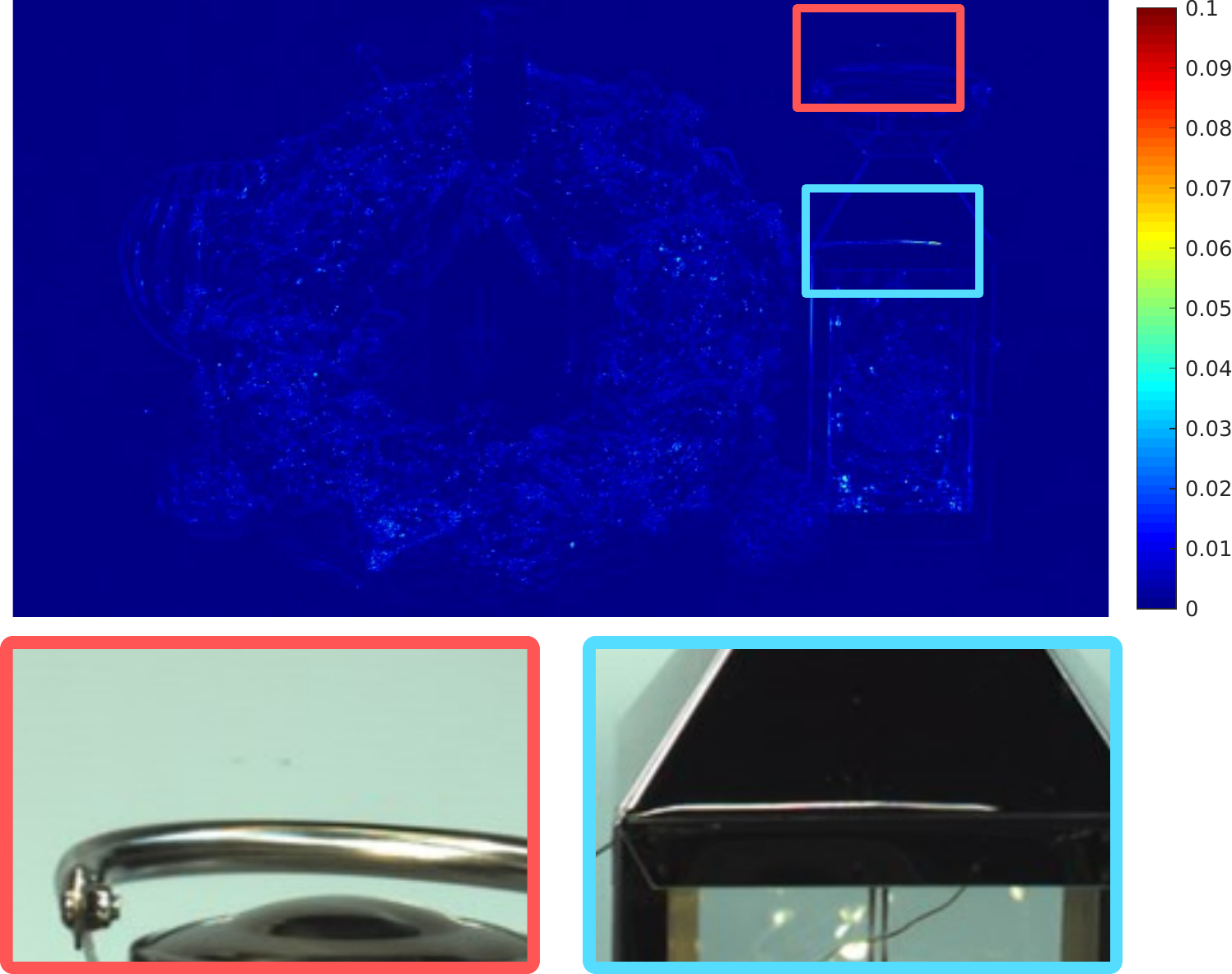}}
  \vspace{-.4em}
  \centerline{\scriptsize(d) DRST (32.708 dB)}\medskip
\end{minipage}
%\hfill

\vspace{-.8em}
\caption{Light field reconstruction results on the evaluation dataset 3.}
\vspace{.4em}
\label{fig:vis3}
\end{figure*}

%\begin{figure*}[t] 
%\begin{minipage}[t]{1.\linewidth}
%  \centering
%  \centerline{\includegraphics[width=.8\textwidth]{Fig/epi_fail/0053.pdf}}
%  \vspace{-.4em}
%  \centerline{\scriptsize(a) The 53-th EPI of $\Psi_3^3$, corresponding to the black dot on the background floor in the red block of \figref{fig:vis3}.}\medskip
%\end{minipage}
%\begin{minipage}[t]{1.\linewidth}
%  \centering
%  \centerline{\includegraphics[width=.8\textwidth]{Fig/epi_fail/0283.pdf}}
%  \vspace{-.4em}
%  \centerline{\scriptsize(b) The 283-th EPI of $\Psi_3^3$, corresponding to the horizontal shiny line in the blue block of \figref{fig:vis3}.}\medskip
%\end{minipage}
%
%\vspace{-.8em}
%\caption{Two ground-truth EPIs of $\Psi_3^3$. Both of them have the same shape, \ie $1280\times 193$ pixels.}
%\vspace{.4em}
%\label{fig:vis4}
%\end{figure*}

%\vspace{-1.em}
\subsection{Results and Analysis}
%\vspace{-.2em}
All the light field reconstruction methods are evaluated quantitatively and qualitatively as bellow. 
\subsubsection{Quantitative evaluation} 	\label{sec:qua}
The minimum and average per-view PSNRs between the reconstructed DSLF $\mathcal{D}$ and ground-truth light field $\Psi$ are utilized to evaluate the light field reconstruction performance. % of different light field reconstruction methods. 
%the proposed and baseline approaches on the above three evaluation datasets.
The quantitative evaluation results of 
DRST and the other three state-of-the-art light field reconstruction methods on the aforementioned three evaluation datasets are presented in \tabref{tab:res1}, \tabref{tab:res2} and \tabref{tab:res3}. %, respectively. 
Looking at the DSLF reconstruction results in \tabref{tab:res1}, it is apparent that DRST outperforms the other three methods \wrt minimal PSNR on all the input SSLFs of the evaluation dataset 1 except for $\mathcal{S}^1_3$.
%which proves the effectiveness of the proposed DRST.
It is noticeable that on $\mathcal{S}^1_1$ and $\mathcal{S}^1_6$, the minimal PSNR results of DRST are 2.964 and 3.825 dB higher than those of the second-best method, \ie ST.
Regarding the average PSNR results, DRST is better than ST on $\mathcal{S}^1_e$, $e\in \{1, 6, 7, 8\}$ and comparable to ST on $\mathcal{S}^1_2$ and $\mathcal{S}^1_4$; 
however, on the rest three input SSLFs, ST achieves better performance than DRST. 
Moreover, both DRST and ST significantly outperform PIASC and SepConv \wrt both minimum and average PSNRs on all the input SSLFs of the evaluation dataset 1. 
The main reason for this is that the light field scenes of the evaluation dataset 1 have repetitive patterns that can hardly be handled by video frame interpolation-based methods, 
since they are incapable of knowing the context information, \ie the moving direction and speed of the virtual camera.  
%More visualization results will be presented in the following section. 
In addition, PIASC and SepConv have almost the same performance on all the input SSLFs of the evaluation dataset 1,
implying that the fine-tuning strategy of PIASC helps little in improving the performance of SepConv on the evaluation dataset 1.

The minimum and average PSNRs of all the light field reconstruction methods on the evaluation dataset 2 are compared in \tabref{tab:res2}.
With regard to minimum PSNR, the proposed DRST performs better than the second-best method, \ie PIASC, on $\mathcal{S}^2_e$, $e\in \{2, 3, 5\}$ and comparably to PIASC on $\mathcal{S}^2_1$, which demonstrates the effectiveness of DRST for DSLF reconstruction in real-world  environments. 
It can also be found that on $\mathcal{S}^2_5$, the minimum PSNR value of DRST is 2.248 dB higher than that of PIASC.
In terms of average PSNR, PIASC achieves the best results among all the four light field reconstruction methods, 
implying that the video frame interpolation-based methods can better handle the DSLF reconstruction for light field scenes without repetitive patterns.
Moreover, ST performs worst among all the light field reconstruction methods \wrt both minimum and average PSNRs. 
Furthermore, the performance of PIASC is slightly better than that of SepConv on all input SSLFs in terms of both minimum and average PSNRs, 
indicating that the the fine-tuning strategy of PIASC is effective in improving the performance of SepConv for the real-world light field scenes of the evaluation dataset 2.

The quantitative results of three light field reconstruction methods on the evaluation dataset 3 are compared in \tabref{tab:res3}.
The results of PIASC are omitted in this table for two reasons: 
(i) PIASC is an enhanced SepConv that is fine-tuned on the ground-truth light fields $\Psi_{1}^3$, $\Psi_{2}^3$ and $\Psi_{3}^3$ of the evaluation dataset 3, 
since these three light fields were the training data provided for the ICME grand challenge; 
(ii) the learning-based SepConv and DRST are neither trained nor fine-turned on the evaluation dataset 3.
It can be seen from the minimum PSNR data in the table that DRST achieves the best performance on three input SSLFs $\mathcal{S}_e^3$, $e\in \{1,4,5\}$.
Besides, DRST is comparable to SepConv on $\mathcal{S}_2^3$.
It can also be found that the minimum PSNR of DRST is 1.432 dB higher than that of SepConv on $\mathcal{S}_1^3$.
As regards average PSNR, DRST performs best on $\mathcal{S}_1^3$ and $\mathcal{S}_4^3$.
Furthermore, in terms of both minimum and average PSNRs, the performance of DRST is better than that of ST,
demonstrating the superiority of DRST over ST. 

\subsubsection{Qualitative evaluation}
The qualitative evaluation results of three light field reconstruction methods on the evaluation dataset 1 are illustrated in \figref{fig:vis1}.
Since SepConv and PIASC perform almost the same as discussed above, the results of SepConv are skipped here. 
The top row exhibits the reconstruction results corresponding to $\mathcal{I}_{93}$ of $\Psi_1^1$.
The checkerboard and Siemens star are chosen as the interesting areas.
As shown in the figure (b), PIASC fails in reconstructing the checkerboard, because this algorithm can hardly exploit the context information;
in other words, 
the moving direction and speed of the checkerboard are unknown to it.  
However, ST and DRST do not have such problem. 
The reconstructed checkerboards are shown in (c) and (d), respectively.
Regarding the Siemens star, all these three methods have small artifacts. 
The bottom row shows the reconstruction results \wrt $\mathcal{I}_{5}$ of $\Psi_6^1$.
The two interesting areas are the Fraunhofer-logo ball and table curtain with repetitive pattern. 
As shown in (f), PIASC fails in reconstructing the table curtain, since the context information is unavailable to it. 
However, ST and DRST overcomes this problem by leveraging the context information implicitly encoded by EPIs. 
Their results are presented in (g) and (h), respectively.
As regards the Fraunhofer logo, ST produces small artifacts when reconstructing the letters, while PIASC and DRST generate visually-correct results. 

The visualized light filed reconstruction results on the evaluation dataset 2 are illustrated in \figref{fig:vis2}.
The results of SepConv are omitted here because the SepConv-based PIASC works slightly better than SepConv on the evaluation dataset 2 as discussed above.
The reconstructed results corresponding to $\mathcal{I}_{58}$ of $\Psi_5^2$ are compared in this figure.
As shown in the red box of (b), PIASC fails in reconstructing the left border correctly. 
However, ST and DRST recover the left border with visually-correct results as shown in (c) and (d), respectively. 
Regarding the reconstruction of the vertical bars close to the right side of the John Deere logo, both ST and DRST have blurry artifacts; nevertheless, PIASC achieves sharp results for the recovery of these vertical bars.

The light field reconstruction results of three different methods on the evaluation dataset 3 are illustrated in \figref{fig:vis3}.
The $\mathcal{I}_{6}$ of $\Psi_3^3$ is chosen to be the reference.
The red and blue blocks in (a) denote two interesting areas.
The red-block interesting area contains the background of the light field scene, which is a flat ground with a black dot. 
The blue-block interesting area has a horizontal shiny line on the metal frame of the lantern.
As shown in (b), SepConv succeeds in reconstructing the black dot and the shiny line with visually-correct results.
However, for both cases, ST and DRST fail in generating visually-correct results.
Specifically, two blurry black dots appear in both (c) and (d),
because the background floor in the real-world light field scenes of the evaluation dataset 3 is out of the disparity range that ST and DRST are designed to handle.
Besides, the shiny line is extended in both (c) and (d),
because both ST and DRST are designed to handle DSLF reconstruction for Lambertian scenes or non-Lambertian scenes consisting of semi-transparent objects only,
while this shiny line is on a non-Lambertian reflection surface.
%The reason can be explained by visualizing the two corresponding ground-truth EPIs in $\Psi_3^3$ displayed in \figref{fig:vis4}.
%As can be seen in \figref{fig:vis4}\,(a), the right thin line corresponds to the black dot in the background region. 

%It should be noted that this line corresponds a disparity value that is much higher than the manually estimated $d_{max}^\mathcal{S}$ in \tabref{tab:res3}.
%In other words, the background floor in the real-world light field scenes of the evaluation dataset 3 are out of the disparity range that ST and DRST can handle, thereby leading to the lower PSNR values of DRST compared to PIASC. 
%In \figref{fig:vis4}\,(b), it can also be found that the border between the shiny line and dark place has a very large disparity value ($\gg d_{max}^\mathcal{S}$ in \tabref{tab:res3}), resulting in the reconstruction errors when using ST or DRST for reconstructing this area. 

\begin{table}[!t]
\begin{center}
\caption{The average computation time (ms) of densely-sampled EPI reconstruction on a color sparsely-sampled EPI $\varepsilon$.} 
\label{tab:time}
\vspace{-.8em}
\scalebox{1.}{
\begin{tabular}{c||c|c|c}
%\hline
\specialrule{.15em}{.0em}{.2em}
  Size of input $\varepsilon$ (pixels) & ST \cite{vagharshakyan2017accelerated} & DRST & Speedup \\[.2ex] 
  \hline
   $1280 \times 13 \times 3$ & 1529.1 & 640.5 & 2.4x \\ 
   $ 960 \times 25 \times 3$ & 3258.9 & 1072.1 & 3.0x \\
   $ 960 \times 13 \times 3$ & 1792.7 & 533.7 & 3.4x \\
   $ 960 \times  7 \times 3$ & 1125.2 & 236.9 & 4.7x \\
%\hline 
\specialrule{.15em}{.2em}{.0em}
\end{tabular}
}
\end{center}
%\vspace{-0.8em}
\end{table}

\subsubsection{Computation time} 
In addition to the above evaluations suggesting that DRST is more effective than ST,  
the computation time of ST and DRST for densely-sampled EPI reconstruction on the input sparsely-sampled EPIs with varying sizes in the aforementioned three evaluation datasets is compared in \tabref{tab:time}. 
As can be seen from this table, the proposed DRST is at least $2.4$ times faster than ST, since DRST performs the shearlet domain transformations for only one iteration. 
Besides, looking at the data of rows one and three, where the angular resolutions of the input sparsely-sampled EPIs are the same, 
DRST achieves a higher speedup over ST for the input EPI with a lower spatial resolution, \ie 960 pixels.
Moreover, it can be seen from the data of rows two, three and four that for the same spatial resolution of the input SSLF, 
the speedup of DRST over ST gets higher when the angular resolution of the input sparsely-sampled EPIs gets smaller, \ie from 25 to 13, then to 7 pixels.
In summary, the value of the speedup of DRST over ST depends on the size of the input sparsely-sampled EPI, \ie the speedup value will be higher if the size of the input sparsely-sampled EPI is smaller.

%\vspace{-.5em}
\section{Conclusion}	\label{sec:conclusion}
%The conclusion goes here.
This paper has presented a novel learning-based method, DRST, for DSLF reconstruction on  SSLFs with disparity ranges up to 16 pixels. 
The proposed DRST takes advantage of a deep CNN, consisting of an encoder-decoder network and a residual learning strategy, to perform sparse regularization in the shearlet transform domain of an input sparsely-sampled EPI,
thereby fulfilling image inpainting on this EPI in its image domain. 
The end-to-end fully convolutional network of DRST is trained on synthetic SSLF data only by leveraging the elaborately-designed masks. 
Experimental results on three different challenging evaluation datasets consisting of real-world light field scenes with varying moderate disparity ranges (8\,-\,16 pixels) show that the learning-based DRST performs better than the non-learning-based ST and comparably to the other state-of-the-art light field reconstruction methods. 
Moreover, DRST is a time-efficient algorithm that is at least 2.4 times faster than ST.

% if have a single appendix:
%\appendix[Proof of the Zonklar Equations]
% or
%\appendix  % for no appendix heading
% do not use \section anymore after \appendix, only \section*
% is possibly needed

% use appendices with more than one appendix
% then use \section to start each appendix
% you must declare a \section before using any
% \subsection or using \label (\appendices by itself
% starts a section numbered zero.)
%

%\appendices
%\section{Proof of the First Zonklar Equation}
%Appendix one text goes here.
%
%% you can choose not to have a title for an appendix
%% if you want by leaving the argument blank
%\section{}
%Appendix two text goes here.
%
%
%% use section* for acknowledgment
%\section*{Acknowledgment}

% Can use something like this to put references on a page
% by themselves when using endfloat and the captionsoff option.
%\newpage

% trigger a \newpage just before the given reference
% number - used to balance the columns on the last page
% adjust value as needed - may need to be readjusted if
% the document is modified later
%\IEEEtriggeratref{8}
% The "triggered" command can be changed if desired:
%\IEEEtriggercmd{\enlargethispage{-5in}}

% references section

% can use a bibliography generated by BibTeX as a .bbl file
% BibTeX documentation can be easily obtained at:
% http://mirror.ctan.org/biblio/bibtex/contrib/doc/
% The IEEEtran BibTeX style support page is at:
% http://www.michaelshell.org/tex/ieeetran/bibtex/
%\small
\bibliographystyle{IEEEtran}
\bibliography{ref}
\end{document}